\RequirePackage[switch, columnwise, running, mathlines, displaymath, mathlines]{lineno}
\RequirePackage{docswitch}
\setjournal{\apj}
\documentclass[iop,revtex4-1]{hackemulateapj}

\usepackage{lineno}
\usepackage{epsfig}
\usepackage{epstopdf}
\usepackage{graphicx}
\usepackage{color,soul}
\usepackage{rotating}
\usepackage{soul}
\usepackage{fancyvrb}
\usepackage{lineno}
\usepackage{amssymb}
\usepackage{amsmath}
\usepackage{xspace}
\usepackage{url}
\usepackage{natbib}
\usepackage{xcolor}
\setstcolor{red}
\usepackage{hyperref}
\hypersetup{
    colorlinks=true,
    linkcolor=teal,
    filecolor=teal,      
    urlcolor=teal,
    citecolor=teal,
}

\usepackage{lsstdesc_macros}

\usepackage{graphicx}
\graphicspath{{./}{./figures/}}

\usepackage{pifont}
\newcommand{\cmark}{\textcolor{teal}{\ding{51}}}
\newcommand{\xmark}{\ding{55}}%

\usepackage{lipsum}
\begin{document}

    \title{Results of the Photometric \textsc{LSST} Astronomical Time-series Classification Challenge (PLA\MakeLowercase{s}T\MakeLowercase{i}CC)}


\newcommand{\URLMODELCALL}{\url{https://plasticcblog.files.wordpress.com/2017/05/noi.pdf}}
\newcommand{\URLSNANA}{\url{http://snana.uchicago.edu}}
\newcommand{\URLMOSFIT}{\url{https://github.com/guillochon/MOSFiT}}
\newcommand{\URLKAGGLE}{\url{https://www.kaggle.com/c/PLAsTiCC-2018}}
\newcommand{\URLMODELS}{\url{https://doi.org/10.5281/zenodo.2612896}}
\newcommand{\URLOPSIM}{\url{http://opsim.lsst.org/runs/minion_1016/data/minion_1016_sqlite.db.gz}}
\newcommand{\URLSextractor}{\url{https://www.astromatic.net/software/sextractor}}
\newcommand{\URLKEYNUM}{\url{https://www.lsst.org/scientists/keynumbers}}
\newcommand{\URLHEALPIX}{\url{http://healpix.sourceforge.net}}
 
\newcommand{\headerOverview}{ Overview of }
\newcommand{\headerDetails}{Technical Details for}
\newcommand{\textEventGenSED}[1]{For event generation, each of the {#1} SED time series was given equal weight.}
\newcommand{\textEventGenLC}[1]{For event generation, each of the {#1} model light curves was given equal weight.}

\newcommand{\HOSTLIB}{{\tt HOSTLIB}}
\newcommand{\SALTII}{{\sc SALT-II}}
\newcommand{\SDSS}{SDSS-II}
\newcommand{\PS}{Pan-STARRS1}
\newcommand{\Diff}{{\tt DiffImg}}
\newcommand{\autoScan}{{\tt autoScan}}
\newcommand{\DESSN}{DES-SN}
\newcommand{\wCDM}{$w$CDM}
\newcommand{\lowz}{low-$z$}
\newcommand{\mosfit}{{\tt MOSFiT}\xspace}
\newcommand{\bands}{$ugrizy$}

\newcommand{\PLASTICC}{Photometric LSST Astronomical Time Series Classification Challenge}
\newcommand{\acro}{{\textsc{PLAsTiCC}}}
\newcommand{\SNANA}{{\tt SNANA}}
\newcommand{\LSST}{Large Synoptic Survey Telescope}
\newcommand{\lsst}{\textsc{LSST}}
\newcommand{\DES}{Dark Energy Survey}
\newcommand{\OPSIM}{{\tt OpSim}}

\newcommand{\Spec}{Spectroscopic}
\newcommand{\spec}{spectroscopic}
\newcommand{\specy}{spectroscopically}
\newcommand{\obs}{observation}
\newcommand{\obss}{observations}
\newcommand{\eff}{efficiency}
\newcommand{\ineff}{inefficiency}
\newcommand{\effs}{efficiencies}
\newcommand{\unc}{uncertainty}
\newcommand{\uncs}{uncertainties}
\newcommand{\Prob}{Probability}
\newcommand{\prob}{probability}
\newcommand{\probs}{probabilities}
\newcommand{\biasCor}{BiasCor}
\newcommand{\con}{contamination}
\newcommand{\Con}{Contamination}
\newcommand{\exgal}{extragalactic}
\newcommand{\Exgal}{Extragalactic}
\newcommand{\Gal}{Galactic}

\newcommand{\MJD}{{\bf\tt MJD}}
\newcommand{\IDEXPT}{{\bf\tt IDEXPT}}
\newcommand{\FLT}{{\bf\tt FLT}}
\newcommand{\GAIN}{{\bf\tt GAIN}}
\newcommand{\RDNOISE}{{\bf\tt RDNOISE}}
\newcommand{\SKYSIG}{{\bf\tt SKYSIG}}
\newcommand{\PSF}{{\bf\tt PSF}}
\newcommand{\ZPTADU}{{\bf\tt ZPTADU}}
\newcommand{\ZPTpe}{{\bf\tt ZPTpe}}
\newcommand{\NEA}{{\bf\tt NEA}}
\newcommand{\PEAKMJD}{{\bf\tt PEAKMJD}}

\newcommand{\ZPHOT}{{\bf\tt ZPHOT}}
\newcommand{\ZPHOTERR}{{\bf\tt ZPHOTERR}}

\newcommand{\Sersic}{S\'ersic}

\newcommand{\LCDM}{\Lambda{\rm CDM}}
\newcommand{\OL}{\Omega_{\Lambda}}
\newcommand{\OM}{\Omega_{\rm M}}
\newcommand{\zcmb}{z_{\rm cmb,true}}
\newcommand{\zcmbObs}{z_{\rm cmb,obs}}
\newcommand{\zhel}{z_{\rm hel,true}}
\newcommand{\zhelObs}{z_{\rm hel,obs}}
\newcommand{\mutrue}{\mu_{\rm true}}
\newcommand{\SNRtrue}{{\rm SNR}_{\rm true}}

\newcommand{\RV}{R_V}
\newcommand{\Msun}{M_{\sun}}
\newcommand\omicron{o}
\newcommand{\DF}{\Delta F} 

\newcommand{\muwCDM}{\mu_{\rm wCDM}}
\newcommand{\muLens}{\mu_{\rm lens}}
\newcommand{\vpec}{v_{\rm pec}}
\newcommand{\vcor}{v_{\rm pec,cor}}
\newcommand{\verr}{v_{\rm pec,err}}
\newcommand{\sigmavpec}{\sigma_{\rm vpec}}
\newcommand{\sigz}{\sigma_{z}}
\newcommand{\dzNoise}{\delta z_{\rm noise}}
\newcommand{\RateUnit}{{\rm yr}^{-1}{\rm Mpc}^{-3}}
\newcommand{\RateUnitGpc}{{\rm yr}^{-1}{\rm Gpc}^{-3}}

\newcommand{\mtrue}{m_{\rm true}}
\newcommand{\Ftrue}{F_{\rm true}}
\newcommand{\FSMP}{F_{\rm SMP}}
\newcommand{\FSIM}{F_{\rm sim}}
\newcommand{\sigF}{\sigma_{F}}
\newcommand{\sigFP}{\sigma_{F}^{\prime}}
\newcommand{\sigFtrue}{\sigma_{\rm Ftrue}}
\newcommand{\sigHOST}{\sigma_{\rm host}}
\newcommand{\sigZPT}{\sigma_{\rm ZPT}}
\newcommand{\sigZERO}{\hat{\sigma}_{0}}
\newcommand{\ERRSCALESIM}{\hat{S}_{\rm sim}}
\newcommand{\ERRSCALESMP}{\hat{S}_{\rm SMP}}
\newcommand{\zSN}{z_{\rm SN}}
\newcommand{\zHOST}{z_{\rm HOST}}
\newcommand{\zSpec}{z_{\rm spec}}
\newcommand{\EFFspec}{E_{\rm spec}}  
\newcommand{\ipeak}{i_{\rm peak}}
\newcommand{\Bpeak}{B_{\rm peak}}
\newcommand{\Trest}{{\rm T}_{\rm rest}}
\newcommand{\Dmb}{\Delta m_b}   

\newcommand{\sigplus}{\sigma_{+}}
\newcommand{\sigminus}{\sigma_{-}}
\newcommand{\cpeak}{\bar{c}}

\newcommand{\SFR}{\rm SFR}
\newcommand{\Ovec}{\vec{\cal O}_{\sigma}}
\newcommand{\mSB}{m_{\rm SB}}
\newcommand{\sigPIPE}{\sigma_{\rm PhotPipe}}
\newcommand{\sigSMP}{\sigma_{\rm SMP}}


\newcommand{\NSIMGEN}{1.1{\times}10^8}
\newcommand{\NSIMTRAIN}{3333}
\newcommand{\NSIMTEST}{3.5{\times}10^6}
\newcommand{\NSIMOBS}{453}  
\newcommand{\NCLASSTOT}{18}
\newcommand{\NCLASSOBS}{14}
\newcommand{\NCLASSOTHER}{4}
\newcommand{\NEXTRAGALMODEL}{15}
\newcommand{\NMOONLSST}{35}
\newcommand{\NMOONZTF}{170}
\newcommand{\NSKYLOCDDF}{133}  
\newcommand{\NSKYLOCWFD}{50,000}  
\newcommand{\NOBSAVGDDF}{330}  
\newcommand{\NOBSAVGWFD}{130}  

\newcommand{\DATECALL}{May 1, 2017 }   
\newcommand{\DATESTART}{Sep 28, 2018}   
\newcommand{\DATESTOP}{Dec 17, 2018}    
\newcommand{\NDAY}{78}                  
\newcommand{\NTEAM}{1,094}
\newcommand{\NPARTICIPANT}{1,384}
\newcommand{\NENTRY}{22,895}
\newcommand{\NMISTAKE}{4}   

\newcommand{\topplace}{Kyle Boone}
\newcommand{\secondplace}{Mike \& Silogram}
\newcommand{\thirdplace}{Major Tom, mamas \& nyanp} 

\newcommand{\classNINENINE}{`Other' class}
\maketitlepre

\begin{abstract}
Next-generation surveys like the Legacy Survey of Space and Time (\textsc{LSST}) on the Vera C. Rubin Observatory will generate orders of magnitude more discoveries of transients and variable stars than previous surveys. 
To prepare for this data deluge, we developed the Photometric \textsc{LSST} Astronomical Time-series Classification Challenge (\textsc{PLAsTiCC}), a competition which aimed to catalyze the development of robust classifiers under \textsc{LSST}-like conditions of a non-representative training set for a large photometric test set of imbalanced classes. 
Over 1,000 teams participated in \textsc{PLAsTiCC}, which was hosted in the Kaggle data science competition platform between \DATESTART~and \DATESTOP, ultimately identifying three winners in February 2019.
Participants produced classifiers employing a diverse set of machine learning techniques including hybrid combinations and ensemble averages of a range of approaches, among them boosted decision trees, neural networks, and multi-layer perceptrons.
The strong performance of the top three classifiers on Type Ia supernovae and kilonovae represent a major improvement over the current state-of-the-art within astronomy.
This paper summarizes the most promising methods and evaluates their results in detail, highlighting future directions both for classifier development and simulation needs for a next generation \textsc{PLAsTiCC} data set.
\end{abstract}

\maketitlepost


\section{Introduction}
\label{sec:intro}

The Legacy Survey of Space and Time \cite[\lsst, ][]{Abell/etal:2009_LSST} of the Vera C. Rubin Observatory will be a survey of the full sky observable from its location in the deserts of northern Chile, with a combination of survey area and depth that is unprecedented. \lsst~will make a complete map of the sky every few days. 
This repeated scanning of the sky will be used to generate difference images formed by subtracting that map from a reference template, leading to millions of ‘alerts’ every night for objects (or detected sources) changing flux. The repeated scanning will also build up light curves (a time series of measured light observed in various filters) for the transients and variable stars \citep{ivezic/etal:2019}, whose brightness changes with time.

This large and heterogeneous data set consisting of signals from different types of astronomical objects will need to be classified into different categories or types of transients and variables from the photometric light curve data. Only a small fraction of objects will be followed up spectroscopically to confirm their types. The need for accurate photometric classification of light curves is therefore a longstanding goal in transient astrophysics. 
Classification challenges have often been designed with a focus on correctly identifying one class from a given data set. 

Previous community efforts to simulate future surveys included the Supernova Photometric Classification Challenge \citep[SNPCC,][]{kessler/etal:2010_SNPhotCC}, which galvanized the astronomy community to 
develop classification methods for three classes of supernovae given a smaller spectroscopic training set. The SNPCC performance metric had a strong emphasis on returning a sample that was both pure (one which had a small number of False Positive classifications) and complete (one which had a large fraction of True Positive classifications), and its performance metric balanced those criteria. Classification challenges must often balance competing goals, with the balance between the two axes often depending on spectroscopic resource efficiency and the brightness limit of any survey, known as the photometric depth. While individual science groups have historically focused on their specific science needs and generated individualized simulations, any classification algorithm that will be run on real and heterogeneous data from \lsst~will need to be robust to that heterogeneity. 

The motivations behind \acro~were somewhat different than those of groups focused on one scientific question. Given the heterogeneous types of objects that we expect \lsst~will provide, and the range of science questions we will want to answer with these data, the \acro~team instead asked the question: \textit{Can you classify a broad range of objects with a good overall/average classification of each type, and identify new objects that are not included in the training set?} The \acro~challenge was presented to the broader public through the Kaggle\footnote{The \acro~challenge is found at \href{https://www.kaggle.com/c/PLAsTiCC-2018}{the PLAsTiCC-2018 Kaggle page}.} platform, a popular site for data challenges and machine learning competitions.  While the participants in \acro~submitted classification methods that were often a combination of different strategies, their metric performance was not tuned to any given type and allowed for comparison of classifiers in cases of degeneracies between classes that had not been explored before. 

The paper is organized as follows:
In Section~\ref{sec:overview}, we present a general \acro\ overview including data simulation, validation, metric and information about the competition.
Section~\ref{sec:results} details the top three submissions and more general aspects of the methods used in the top classifiers.
We study alternative methods of exploring the performance of the classifiers in Section~\ref{sec:alternative_metrics}, and summarize the challenges presented by the data in Section~\ref{sec:challenges}. 
We conclude with key insights from this challenge in Section~\ref{sec:discussion/conclusion}.

\section{Overview of \acro}\label{sec:overview}

\subsection{Data simulation}

\begin{figure*}[htbp!]
\begin{center}
\begin{tabular}{ll}
\includegraphics[scale=0.4]{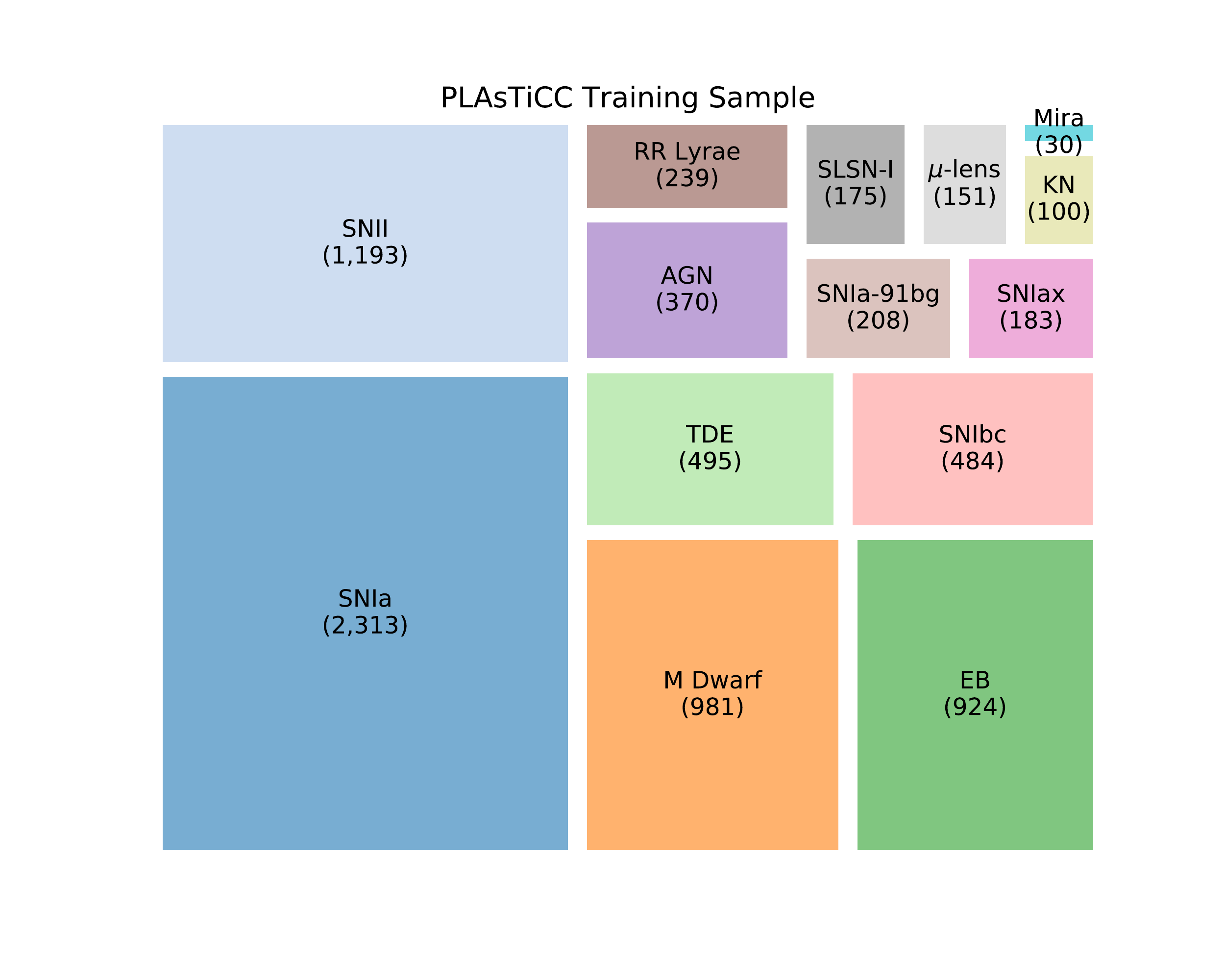}&\includegraphics[scale=0.4]{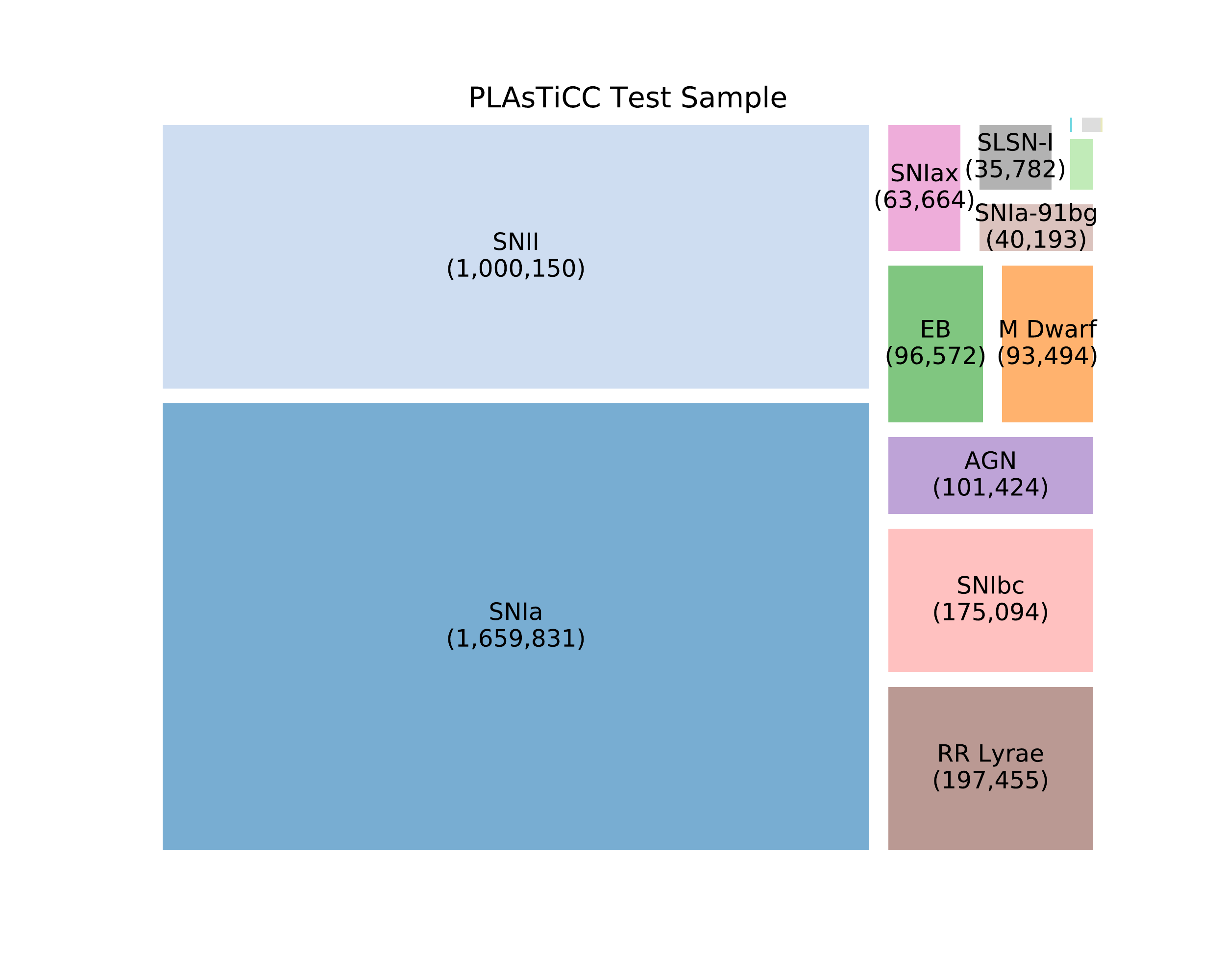}
\end{tabular} 
    \caption{Relative numbers of objects of each class between the training (left) and test (right) data of \acro. The size of the boxes are proportional to the relative numbers in each set and the absolute numbers are in parentheses. The simulated $\simeq 8000$ objects in the training set were distributed across the data classes in a different ratio compared to the test data of over 3.5 M objects.}
    \label{fig:train_to_test}
\end{center}
\end{figure*}

The \acro~data were simulated to mimic three years of an \lsst-type survey (which is slated to run for a decade) using the SuperNova ANAlysis package \citep[\SNANA][]{snana}. In May 2017, an initial call was put out to the astronomy community to develop models of extragalactic transients, Galactic variables and novel astronomical transients. In response to the call, 18 different models were generated according to estimated rates and distributions on the sky. The models included 
\begin{enumerate}
    \item Extragalactic models -- including exploding supernovae (SNe) which have large variations in peak brightness, kilonova (KN) models which are the electromagnetic output from two colliding neutron stars, the emission from the tidal disruption events (TDE) of stars due to the supermassive black hole at the center of a galaxy, and emission from gas falling onto the black hole creating an active galactic nucleus (AGN). The supernova models represent a wide variety of scenarios, from the explosion of stars at the end of their lives (SNII, SNIbc), exploding white dwarf progenitors (the SNIa models including SNIa, SNIa-91bg and SNIax) and the super-luminous supernovae (SLSN).  
    \item Galactic models -- which include variable stars and recurring transients such as Mira variables, RR Lyrae and eclipsing binaries (EB), and non-recurring models such as the single microlensing events ($\mu$lens-Single) and the M-dwarf flares.
    \item `Novel' objects -- Pair Instability Supernovae (PISN), Calcium-rich transients (CaRTs), Intermediate Luminosity Optical Transients (ILOTs), binary micro-lensing events ($\mu$lens-Binary).
\end{enumerate}
The simulation generated 100 million sources, and a subset of 3.5 million were predicted 
to be detected by \lsst, and thus useful for classification and analysis.
These data were provided to participants in \acro~as 3.5 million light curves in the $ugrizY$ filters
containing over 453 million observations. These data formed the `test' sample that 
\acro\ participants were asked to classify. 
The training set consisted of 8000 objects constructed as a mock-spectroscopic subset of the full data, meant to mimic the available data from current and near-term spectroscopic surveys that will be available at the start of \lsst~science operations.
The differences in numbers and representation between the training and test data sets can be seen in Figure~\ref{fig:train_to_test}.

The \acro~data were simulated to mimic future \lsst~observations in both the Deep Drilling Fields (DDF), which are small patches of the sky that will be sampled often to achieve enhanced depth, and a Wide-Fast-Deep (WFD) survey that covers a larger part of the sky 
(at $\simeq 18,000$ square degrees it is almost 400 times the DDF area)
that will be observed less frequently. Compared to DDF, WFD will include many more objects but have lower signal-to-noise ratios. While the \acro~simulation was performed for one particular survey configuration that is not optimal for transients, classification performance can be used to evaluate survey design for \lsst. We leave this to future work.
The extragalactic models were simulated from the \acro\ models as described in \cite{kessler/etal:2019_PLAsTiCC} (see Figure 13 for the full simulation procedure) and \cite{kessler/etal:2019_DES}, by generating a spectral energy distribution (SED) at each epoch for the source, simulating a cosmological distance and peculiar velocity for the source, computing the lensing seen by that source, redshifting the SED, and accounting for Galactic extinction. Each extragalactic object was simulated with a unique redshift, host-galaxy extinction, right ascension and declination, and cadence. The SED is integrated in each filter to produce a light curve for the object. The Galactic models are instead defined by 4-year time sequence of true magnitudes in the $ugrizY$ filter bands, which are then sampled to form a light curve. The rates for the different objects were supplied by the modelers. For the Galactic models, these rates included a dependence on Galactic coordinates, while the extragalactic rates were assumed to be isotropic. Galactic and AGN models were simulated with unique initial phase, reference-image flux, right ascension and declination and cadence.

Once the `pure' source model was obtained, it was combined with a noise model specific to the observational conditions of \lsst~\citep{biswas/opsim}, including the cadence information, zero points, sky noise, and point spread function (PSF) \citep{snana}. The objects have to be `detected' in order to be included in \acro. We required that each source be detected twice at roughly $5\sigma$ in the difference image between two nights -- so sources must be bright and vary in brightness to be valid \acro~objects. For detected events, a training subset was tagged as spectroscopically identified, and another subset was tagged as having a spectroscopically measured redshift of the host galaxy.
 
\lsst~may well discover new transients that are not captured by current models, and it may be sensitive to theorized objects that have not yet been observed due to the limitations of current surveys. Thus the training data will not be fully representative of the larger test data that will probe fainter depths and larger redshifts than the training data. To capture this non-representativity, four of the models simulated in the \acro~test data, namely the PISN, CaRTs, ILOT, $\mu$lens-Binary were  \textit{not} provided in the training set \citep[see][for individual model references]{kessler/etal:2019_PLAsTiCC}. As part of the challenge, participants were asked to classify objects in the test set into a different class of previously unknown objects described also as the \classNINENINE. Participants were not given any information about the \classNINENINE~or the composite nature of this class, which is discussed in more detail in Section~\ref{sec:results}. 
The simulation procedure and the models involved are discussed in extensive detail in \cite{kessler/etal:2019_PLAsTiCC}. The models are provided for use by the community \citep{plasticc_modelers_2019_2612896}.

\subsection{Data validation}
\label{sec:validation}

\begin{figure*}
    \centering
    \includegraphics[width=0.9\textwidth]{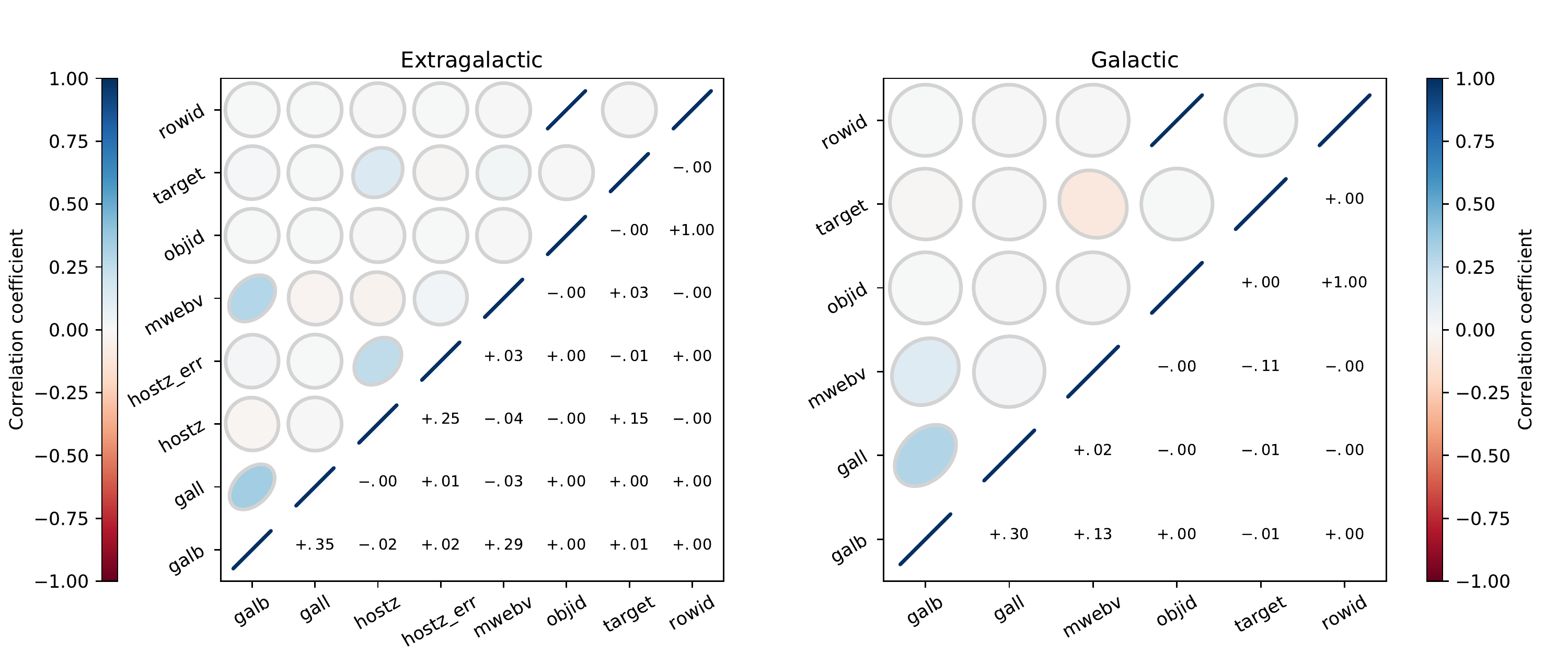}
    \caption{Correlation tests for the simulated WFD \acro~data. We used this test to detect any spurious correlations between object ID and type and other properties provided to the contestants. We disambiguate Galactic and extragalactic sources and consider the WFD and DDF data separately as there are physical correlations that are introduced when aggregating these groups e.g. all Galactic variables will have a host redshift of $z=0$, and all sources within the 4 DDF fields will show strongly-correlated values of the Milky Way reddening. }
    \label{fig:validation_corrs}
\end{figure*}

One of the most critical parts of developing \acro~was validating the data, to ensure that all light curves had been simulated without irregularities or too large/small errors, and that unphysical correlations (e.g., catalog ID number vs. declination) were not present. In addition, we validated the physical models that were input to the simulations. We also validated and improved the \SNANA~simulation code. We list the validations performed on the simulations below.

\textbf{Unphysical artifacts --}
When processing light curves and metadata (additional information on the objects including their sky position, redshift etc.) in order to classify objects, unphysical artifacts or missing light curve data can lead to leakage. As one example, modelers typically provided SED information starting at $1000~\mathrm{\AA}$ or $2000~\mathrm{\AA}$, which will cause the $u$-band (with central wavelength around $3000~\mathrm{\AA}$) to drop out at extreme redshift ($z\simeq3$). To avoid UV drop-outs in the simulation, we extrapolated the UV flux to $500~\mathrm{\AA}$ to ensure that all $u$-band observation were included \citep{kessler/etal:2019_PLAsTiCC}. If left unchecked, classifiers might use this zero flux as a feature, however this is unphysical and should be ignored. If we were unable to debug or account for an artifact, the affected data were removed.

\textbf{Distribution tests --}
The \acro~data were simulated from model objects at a given rate and from SEDs that were redshifted, and then `observed' in the \lsst~bands with suitable noise properties and observing conditions. The first step in the validation procedure was to perform tests of the distribution properties of the objects to ensure that they matched those of known objects for models based on observations or expected distributions for theoretically proposed models. We performed checks of the distributions of the maximum and minimum flux for the objects, their redshift distribution, and rates. These tests caught a host of minor issues in reported rates; in many cases the model assumptions were revised for inclusion in the final suite of models. The real data from \lsst~will not be validated in exactly the same way given that much of the initial classification will  proceed from the alert stream directly; however quality cuts on the data will ensure data quality for any light curves used for a specific scientific question.

\textbf{Light curve tests --}
Once the average properties of the population were inspected, individual light curves were plotted from the simulated populations. These tests were particularly relevant for models derived from objects that had previously been observed (rather than objects simulated from purely theoretical models). The light curves were compared to databases of known examples to diagnose model inconsistencies and shape discontinuities. This validation process led to decisions on how the \acro~simulation would handle saturation in the \lsst~bands for objects brighter than 16th magnitude. However, we did not want to introduce saturation issues in the challenge, and instead simulated a saturation level at 12th mag, or 4 mag
brighter than nominal. The small number of saturated observations that remained were removed.

Prior to generating the \acro~sample, we simulated `perfect' examples of each model type at high cadence and with no noise. These perfect light curves were compared with the input observations used to generate the model, and then with the results of small runs of the simulations in both the DDF (which would probe higher redshift with higher signal-to-noise) and the WFD, to test the rates of objects more completely and whether or not there were objects that would not be detected given the cadence and brightness specifications of the survey. 
Perfect simulations were used to validate the models while the WFD and DDF light curves were used to validate the realistic simulations.

\textbf{Specialized model tests --}
Depending on the type of transient or variable, specialized tests on object properties were carried out for individual models. For example the period-luminosity relationship was tested for variable objects like the RR-Lyrae. These tests were typically performed with input from the model proposers directly, to ensure that the expected model relationships remained intact in the \acro~simulations.

\textbf{Checks on metadata -- }
Once the models were verified for correctness, the additional data about the objects, or metadata (the redshifts, sky position, Galactic coordinates, extinction) were analyzed to ensure that the data are simultaneously useful and exhaustive without rendering the challenge trivial. This involved iterating on the form and description of the metadata, knowing that the user interface/workbook provided to initiate the challenge would be the place where many participants got their information. 
The metadata were also checked for consistency by, for example, plotting RA and Dec per target. 

\textbf{Correlations between object variables --}
Related to checks of the metadata, we tested for spurious or obvious correlations between object metadata and classification type. A trivial example would be to ensure that the object ID was in fact randomly generated not only within a particular type but across various types. If the objects are simulated in order then there would be a type-ID correlation which would render the classification challenge trivial. Some astrophysical correlations do exist, for example the nature of the objects (Galactic vs. extragalactic) is correlated with the redshift. A plot of the correlation between training set metadata parameters is shown in Figure~\ref{fig:validation_corrs}, where the simulated WFD sample has been split into Galactic and extragalactic objects. There are correlations in the metadata, but they are expected. For example, the redshift is correlated with the redshift error and the target is associated with the dust reddenning of the Milky Way (Galactic objects will have more dust than extragalactic). We also do not see unphysical correlations such as with object ID or row ID.

\textbf{Simple classification of objects --}
An additional test of the robustness of the training set was to train a `simple' random forest classification algorithm \citep{narayan/etal:2018, random_forest} on the \acro~data to ensure that the problem was not trivially solvable, and also provided a rough benchmark for comparing results throughout the challenge. The 11 extracted features per passband are described in Table 2 of \cite{narayan/etal:2018}, and included the autocorrelation length, the kurtosis and skewness of the magnitude distribution. The classifier was run on the DDF and the WFD separately during the validation phase and as such was not run on the complete data set after production. 
Given that some rare objects such as the kilonovae were not found in the DDF, a direct comparison of the performance of the simple classifier to the competition entries is less straightforward; however the simple classifier was merely testing for obvious problems in the simulated sample. In addition, the simple classifier was not run on the \classNINENINE~objects, as testing anomaly detection was not part of the validation process. This classification testing (and the other model validation) was only performed by members of the \acro~team who had agreed \textit{not to submit an entry to the challenge, and not to help challenge participants}. Each model type was validated by a subset of the \acro~development team and was re-validated for each new version of the simulated data.

\subsection{The \acro~metric}

The choice of metric for \acro~was motivated by two main drivers: the importance of classification uncertainties and the lack of a single, unifying science goal.
Unlike its predecessor, the SNPCC, which was motivated by SNe Ia cosmology, \acro~ was motivated by dozens of science drivers from Galactic population statistics to gravitational wave electromagnetic counterparts.
The metric was designed to be sufficiently generic so that it would disfavor classifiers that neglected any classes, especially those that are rare in the Universe.

Because the data from \lsst~is anticipated to be inherently noisy, the \acro~metric did not reduce the classification posteriors to deterministic class estimates, allowing it to include classification uncertainty. In contrast to traditional classification tasks that request a single estimated class $\hat{m}_{n}$ for each classified object $n$, \acro~ required participants to submit tables of classification posterior probabilities $0 \leq p(m \mid d_{n}) \leq 1$ where object $n = 1, \dots, N$ belongs to class $m = 1, \dots, M$, given the light curve data $d_n$. 
The \acro~metric was a weighted log-loss
\begin{eqnarray}
  \label{eq:logloss}
  L &\equiv & -\sum_{m=1}^{M} w_{m} \sum_{n=1}^{N_{m}} \tau_{n, m} \ln[p(m \mid d_{n})],
\end{eqnarray}
where
\begin{eqnarray}
  \label{eq:indicator}
  \tau_{n, m} &\equiv& 
  \begin{cases}
  0 & m_{n} \neq m \\
  1 & m_{n} = m
  \end{cases}
\end{eqnarray}
is an indicator variable for the true class.
This particular metric strongly disfavors assigning very low probabilities to the classes that the classifier thinks are not the correct one, implying that very low entropy classifications (i.e. with lots of zeros/small values in the vector of class probabilities) are disfavored.

\begin{table}[htbp!]
\begin{tabular}{c|c|c}
   Class Name  & Class number & Weight \\
   \hline
Point source $\mu$-lensing     & 6 & 2\\
Tidal disruption event (TDE)     & 15 & 2\\
Eclipsing binary event (EBE)     & 16 & 1\\
Core-collapse supernova  Type II (SNII)     & 42 & 1\\
Supernova Type Ia-x  (SNIax) & 52 & 1\\
Mira Variable    & 53 & 1\\
 Core-collapse Supernova  Type Ibc   (SNIbc) & 62 &1 \\
Kilonova (KN)       & 64 &2 \\
M-dwarf     & 65 &1\\
Supernova  Type Ia-91bg (SNIa-91bg)    & 67 & 1\\
Active galactic nucleus (AGN)     & 88 & 1\\
Supernova  Type Ia  (SNIa)   & 90 &1 \\
RR Lyrae     & 92 & 1\\
 Super Luminous Supernova (SLSN)    & 95 & 2 \\
\classNINENINE     & 99 &2 \\
\end{tabular}
\caption{Table of weights and target numbers for the various classification types included in \acro. Rare or interesting objects were up-weighted to increase their contribution to the total metric. \label{tab:weights}}
\end{table}

The weights $w_{m} = J / N_{M}$, where $J=1$ for most classes, but $J=2$ for rare objects, as shown in Table~\ref{tab:weights}. 
These weights were chosen by the \acro~team to discourage participants from classifying all objects as the most populous class, ensuring that appropriate attention was directed to the classification of rare classes that otherwise would not have much influence on a metric that gave equal weight to all objects. 
A detailed description of the ~\acro~metric may be found in \cite{malz/etal:2019_metric}, along with an investigation of its assessment of archetypal classification pathologies.

\begin{figure}
    \centering
    \includegraphics[width=0.95\columnwidth]{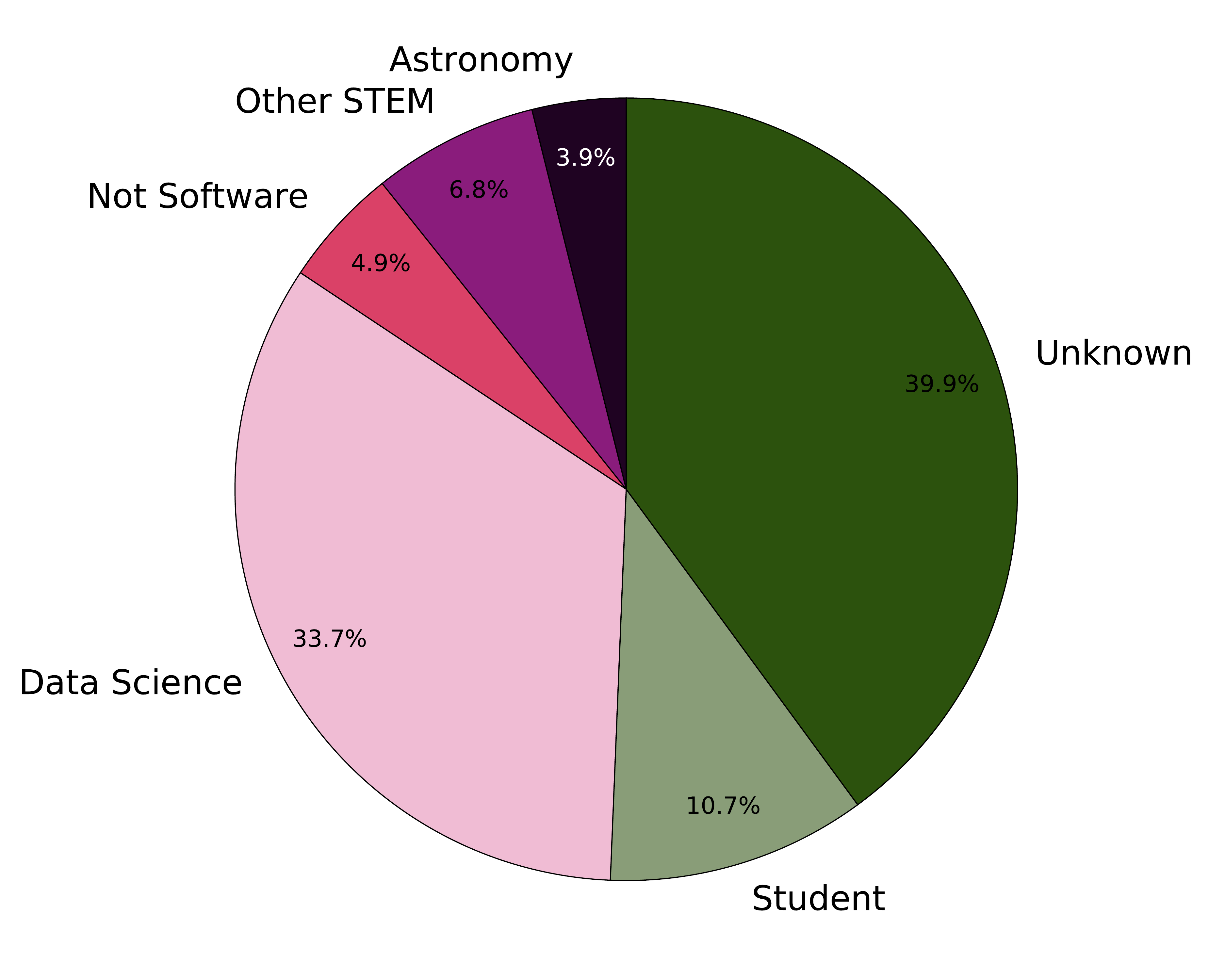}
    \caption{Demographic information for entrants to \acro. Only a small fraction of 1315 participants with known demographics were part of the astronomy community. Many of the participants have some data science background. These data were determined from the Kaggle entries or GitHub user pages of the entrants, future studies will include self-identification information from participants.
    \label{fig:pie}}
\end{figure}

\subsection{The \acro~competition}
\label{sec:submissions}

\begin{figure*}[htbp!]
\begin{center}
\begin{tabular}{l}
\includegraphics[scale=0.6]{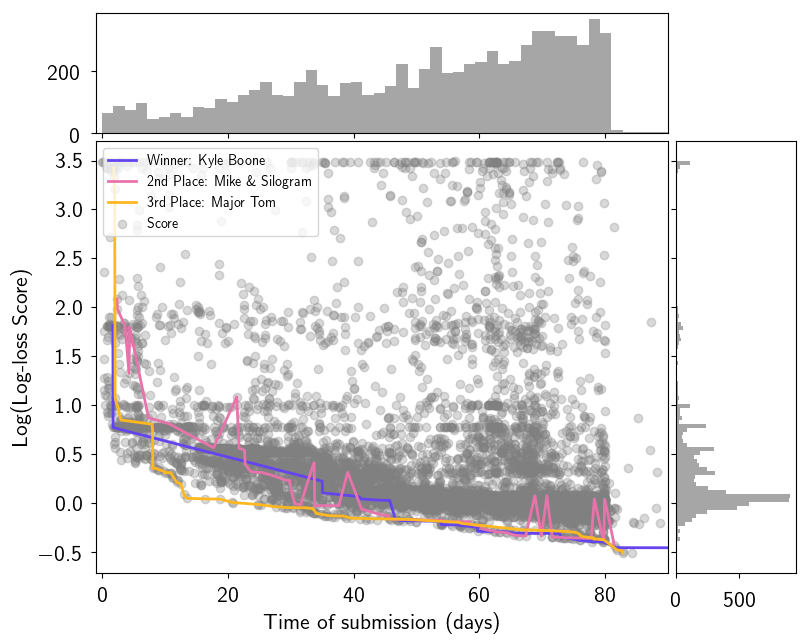}
\end{tabular} 
\caption{The distribution of the submissions to \acro~and the log of the \acro~metric score  as a function of time since the start of the challenge relative to the start of the challenge. The three colored lines indicate the performance of the three prize-winning entries. As the challenge neared completion, the scores started to asymptote to metric scores of near unity. Note that the competition remains `live' to this day, hence the few submissions received after the closing deadline.
\label{fig:submissions}}
\end{center}
\end{figure*}

\acro~ran from \DATESTART~until \DATESTOP, and received entries from 1094 teams around the world from a variety of different groups. The demographic information of the 1315 participants are summarized in Figure~\ref{fig:pie} and was determined from the Kaggle or GitHub profiles of the participants in each team. Some of the participants have subsequently deleted their accounts, or do not have complete profiles, and so Figure~\ref{fig:pie} provides an estimate of the group demographics. While a substantial fraction of participants have a background in data science/software, only a fraction of entrants had formal astronomy training.
A reasonable subset of the \acro~participants were students. These included a group from the University of Warwick in the United Kingdom who entered \acro~as part of a class project.

\acro~led to publications by participating groups, including presentations of the winning \texttt{avocado} algorithm \citep{boone:2019} and that of the group `Day meets night' \citep{gabruseva/etal:2020}. The data set itself has also been used for several publications on classification methods and anomaly detection approaches \citep{ishida/etal:2020,soraisam/etal:2020,sravan/etal:2020,dobryakov/etal:2020,chaini/etal:2020}.

Though participants submitted their classifications on the full \acro~test data set, the metric was evaluated on representative subsets whose distinctions were not made known to participants. 
The metric computed on one third of the full data set was posted to a ``public leader board,'' whereas the metric evaluated on the remaining two-thirds of the test data was the basis for the official ranking Kaggle used to award the cash prize. 
This procedure could be applied to subsets of the data to determine a classification `error', however Kaggle did not include any such error estimation in their application of the \acro~metric.

This approach of keeping data hidden from participants is in place to discourage `probing the leader board' (hereafter LB probing), wherein one's position on the LB is used as a metric on its own to tweak a submitted algorithm until it is tuned to the LB performance. While this will lead to improved performance in one particular challenge/realization of the data, this approach does not yield algorithms robust to changes in data simulation, or more general intuition-building about the problem.

The distribution of submissions over time and their corresponding metric values are shown in Figure~\ref{fig:submissions}.
Interest in the challenge grew over time as more participants joined. 
Kaggle restricts participating teams to a maximum of five submissions per day, a limitation that proved critical near the end of the challenge, when participants had to be selective in making submissions that would best inform final adjustments to their classifiers. 

The performance of the three winning solutions are highlighted in the plot as colored lines. The horizontal clustering of the \acro~metric near scores of $\log[\mathrm{weighted\ log\textnormal{-}loss\ score}] = 1, 1.6$ (where the log-loss score is given in Equation~\ref{eq:logloss}) can be seen in the 1-dimensional histogram of scores on the right-hand panel of Figure~\ref{fig:submissions} and could be a result of over-fitting (`playing the leader board'), particularly where weighted combinations of classification probabilities for some objects are used to determine the probabilities of other objects (e.g. the \classNINENINE~objects).

\section{Solutions to \acro}
\label{sec:results}

\begin{figure*}[htbp!]
\begin{center}
\begin{tabular}{l}
\includegraphics[width=0.8\textwidth]{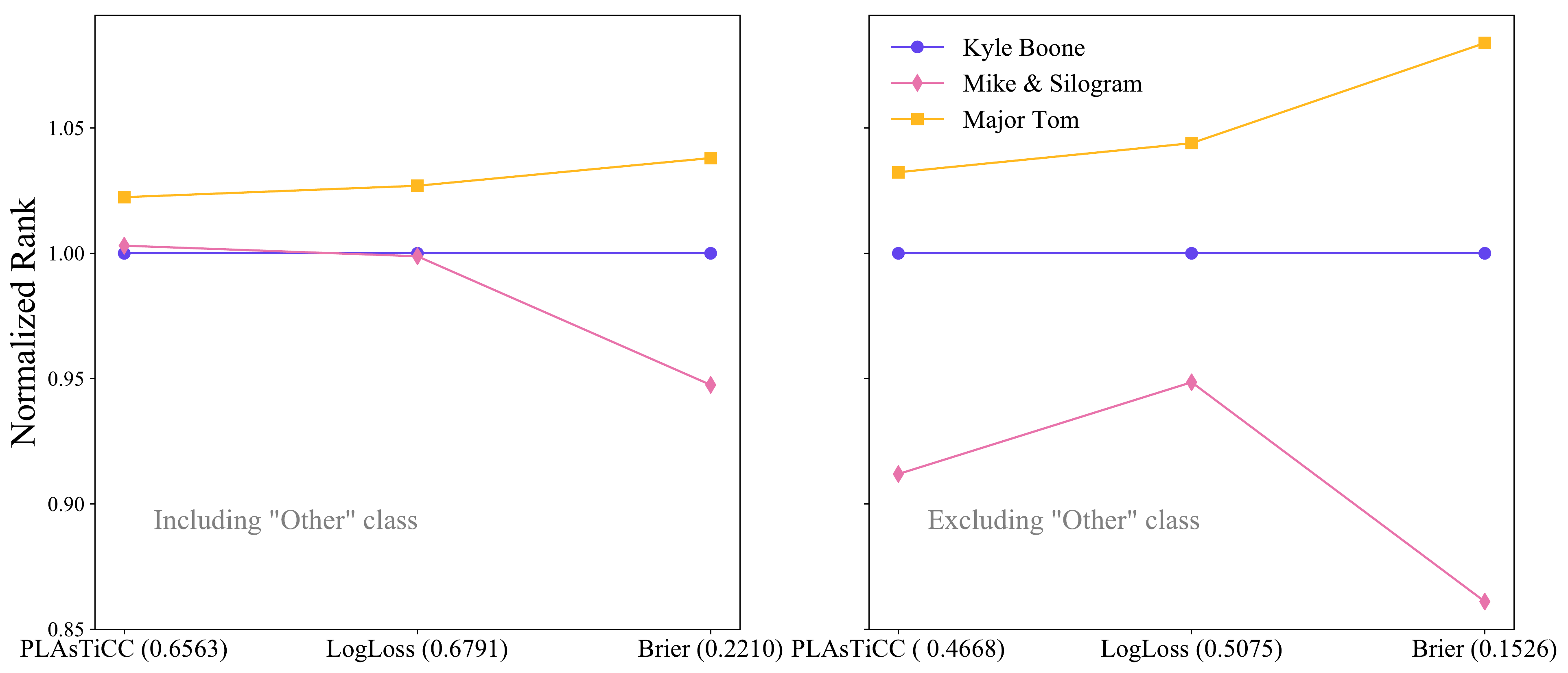}
\end{tabular} 
\caption{The performance of the top three entrants to the \acro~across three different metrics. In each case a lower score indicates better performance of the classifier. The left panel shows the metrics for the full test data set including the unseen \classNINENINE~objects. All classifiers were asked to return probabilities for the \classNINENINE~objects. The right panel shows the scores on the test data set excluding the \classNINENINE~objects. The scores have been normalized to the score of \topplace, indicated in parenthesis on the $x$-axis label. This plot illustrates how close the first and second place finish was, and the importance of the unknown objects on the final classifier performance. In the case where the \classNINENINE~objects were not included, the first and second place positions would have been reversed. 
\label{fig:metrics}}
\end{center}
\end{figure*}

\acro~ended on December 17, 2018. 
The competition was won by \topplace~which is described in \cite{boone:2019}, with the second place prize awarded to \secondplace\footnote{The second place solution is described in a \href{https://www.kaggle.com/c/PLAsTiCC-2018/discussion/75059}{Kaggle post}.} and third place was awarded to ``\thirdplace''\footnote{The third place solution is described in two separate posts, one describing the \href{https://www.kaggle.com/c/PLAsTiCC-2018/discussion/75116}{CNN} and another describing the \href{https://www.kaggle.com/c/PLAsTiCC-2018/discussion/75131}{model including code repositories}.}.
The top three teams were selected to present their methods at a meeting with the \acro~team and the staff at Kaggle.  

Figure~\ref{fig:metrics} shows the performance of the top three entries according to the metrics considered in \citet{malz/etal:2019_metric}, namely the weighted log-loss used in \acro, an un-weighted log-loss, and the Brier score \citep[][equivalent to the mean square error of the prediciton]{brier1950}, normalized to the \topplace~score. The left panel shows the metrics evaluated on the full test set (rather than the 1/3 subset that formed the basis of the \acro~competition results), while the right panel performance when the \classNINENINE~objects were excluded from the test data. The scores of the top three entries tighten in that case and the top two places are reversed, highlighting the importance of returning a classification probability for objects whose class was not present in the training set.
All three top finishers were very close in score throughout the competition in both the public and the private LBs. We summarize their solutions and their relative performance in the subsections that follow.

\begin{table*}[htbp!]
\begin{center}
    \begin{tabular}{rcccccccc}
    \hline
    \hline
         Final & Name& Public & Private & Class imbalance  & Non-representativity & \classNINENINE~& Ensemble & Light curve  \\
    Rank & & score & score &  mitigation &  mitigation & &  & fit \\
    \hline
    1 & Kyle Boone & 0.6706 & 0.6850 & Light curve & Data degradation & Weighted Average  & No  & GP \\ 
      &  &  &  &  augmentation &  &   &   &  \\
    \hline
    2 & \secondplace & 0.6937 & 0.6993 & - & Model spec-$z$,  & Probability &  Yes &  - \\
     &   &  &  &  & Data degradation  & Scaling &  & \\
    \hline
    3 & Major Tom, mamas & 0.6804 & 0.7002 & - & Model spec-$z$,  & Tuned &  Yes & SALT2 \\ 
     & $\&$ nyanp &  &  &  & Data degradation  & Combination & & \\
    \hline
    4 & Ahmet Erdem & 0.6913 & 0.7042 & - & Model spec-$z$ & Weight based on &  Yes & Bazin \\ 
     &   &  &  &  &  & top prediction &  & \\
    \hline
    5 & SKZ Lost in & 0.7397 & 0.7523 & Flux error & Removed photo-$z$ & Probability &  Yes & - \\ 
    & Translation  &  &  & augmentation &  & Scaling &  & \\
    \hline
    6 & Stefan Stefanov & 0.7933 & 0.8017 & Flux augmentation & Dropped light curve  & - &  Yes & - \\
     &   &  &  &  &points, photo-$z$  &  &  & \\
        &   &  &  &  &augmentation  &  &  & \\
    \hline
    8 & rapids.ai & 0.7922 & 0.8091 & Pseudo label & -  & - &  Yes & - \\
    &   &  &  & on test data &  &  &  & - \\
    \hline
    9 & Three Musketeers & 0.7921 & 0.8131 & - & Dropped light curve p & Probability &  Yes & - \\ 
    &   &  &  & &oints  & Scaling &  & \\
    \hline
    11 & Simon Chen & 0.7948 & 0.8225 & - & - & - & No & SALT2 \\ 
    \hline
    12 & Go Spartans! & 0.8112 & 0.8265 & - & - & Probability & Yes & Bazin \\ %
    &   &  &  & &  & Scaling &  & \\
    \hline
    \end{tabular}
    \end{center}
    \caption{Methods employed by classifiers in \acro. The scores are taken from the Kaggle site, and illustrate how many methods were tuned to the public data/scores, and performed worse on the private data. The class imbalance techniques were used to address the fact that the different classes were present in very different numbers in the data set. Techniques include augmenting the light curve or fluxes, and adding approximate/pseudo `labels' to the test data to supplement the training data. The non-representativity techniques modified data to account for the differences in the test and training data (e.g. that the test data were noisier and more sparse) by either degrading the data (error bars), removing light curve points and modelling the spectroscopic redshift (spec-$z$) from the photometric redshifts (photo-$z$). The light curve fitting techniques included Gaussian Process (GP) interpolation and template fitting \citep[using the ][light curve fits]{bazin/etal:2012,salt2}. As discussed in Section~\ref{sec:methods}, the entrants used a range of methods listed here to derive a probability for the \classNINENINE~objects, and then probed the LB to optimize their procedure. In all columns above, if there is no entry for a given column, it indicates either that the model did not address the column or that the information was not provided in the write-ups presented on Kaggle.}
    \label{tab:methods}
\end{table*}

\begin{table*}
\begin{center}
\begin{tabular}{ccccccccc}
    \hline
    \hline
    Name & \multicolumn{3}{c}{Boosted Decision Trees} & & \multicolumn{4}{c}{Neural Nets} \\
    \cline{2-4}
    \cline{6-9}
 & LightGBM & CatBoost &  XGBoost &  & NN &  CNN & RNN & MLP \\
    \hline
    \topplace & \cmark & \xmark & \xmark & & \xmark & \xmark & \xmark & \xmark  \\
    \secondplace & \cmark & \xmark & \xmark & & \xmark & \xmark & \cmark & \xmark \\
    \thirdplace & \cmark & \cmark & \xmark & & \xmark  & \cmark & \xmark & \xmark   \\
    Ahmet Erdem & \cmark & \xmark & \xmark & &  \cmark & \xmark & \xmark & \xmark   \\
    SKZ Lost in Translation & \cmark & \xmark & \xmark & & \xmark & \xmark & \cmark & \cmark  \\
    Stefan Stefanov & \xmark & \xmark & \xmark & & \cmark & \xmark & \xmark & \xmark  \\
    rapids.ai & \cmark & \xmark & \xmark & & \xmark & \xmark & \cmark & \cmark  \\
    Three Musketeers & \cmark & \cmark & \cmark & & \xmark & \cmark & \xmark & \xmark \\
    Simon Chen & \cmark & \xmark & \xmark & & \xmark & \xmark & \xmark & \xmark  \\
    Go Spartans! & \cmark & \xmark & \cmark & & \xmark & \xmark & \xmark & \xmark \\
    \hline
    \end{tabular}
\caption{Classification algorithms of the top entrants to \acro, sorted into common approaches. A key result of the challenge was the observation of how different solutions combined various machine learning techniques. The various methods are summarized in Table~\ref{tab:acronyms}.
\label{tab:algorithms}}
\end{center}
\end{table*}

\subsection{Summary of methods used}
\label{sec:methods}

The Kaggle discussion board was an excellent forum for the dissemination and discussion of methods used to provide solutions to \acro. The entrants from the top 12\footnote{The 7th and 10th place entries did not provide their solutions on the Kaggle forums so are not included here.} entrees discussed their solutions on the forum, and here we briefly summarize the elements used in their solutions in Table~\ref{tab:methods} and Table~\ref{tab:algorithms} shows the classification algorithms that were implemented.

\textbf{Combining multiple approaches -- } The unifying characteristic of the solutions is the fact that they employed a mixture of classification approaches from neural networks to gradient boosting. The entrants often tried different combinations of approaches and adjusted their mixture within different iterations, effectively playing multiple classification methods `against' each other internally before submitting their solutions at each step. We refer to this as `Ensemble' in Table~\ref{tab:methods}. Common among most solutions was the use of the LightGBM \cite{lightgbm}. Approaches varied when it came to feature selection. Some groups built on previous attempts by performing template light curve fits to extract features, using known fitting software such as the {\texttt SALT2} package \citep{salt2} or the \cite{bazin/etal:2012} model. Some used the more time-consuming but less model-dependent Gaussian process (GP) fitting. Originally developed in the context of mining surveying \citep{krige:1951}, GP interpolation relies on the assumption that the interpolated values can be modeled by a Gaussian process, or one in which the process is defined by a prior covariance, or kernel. This interpolation allows one to easily map not only the interpolated values, but the uncertainty band around the interpolation without assuming any physical relationship between the variables. This makes it well-suited to light curve fitting.

\begin{table*}[]
    \begin{center}
    \begin{tabular}{p{0.2\textwidth}|p{0.5\textwidth}|p{0.2\textwidth}}
    Method & Description & Reference \\
    \hline
        \hline
        Random Forest & Also known as random decision trees, random forests are a &\\&method of classification that constructs many decision trees&\\& between different classes. The mode of multiple trees &\\&is computed to obtain a final classification. & \cite{random_forest}\\
        \hline
        Gradient Boosting & Gradient boosting produces a combined model from &\\& an ensemble of prediction models (often but not &\\& always comprised of decision trees).
        Each new tree is &\\& fit on a modified version of the original data set, and the weights of &\\&subsequent trees are adjusted according to which objects were difficult &\\& to classify by previous trees (upweighting difficult classifications).& \cite{gradient}\\
        \hline 
        Light Gradient  & LightGBM is a gradient boosting framework originally &\\Boosting Machine& developed by Microsoft, but now open source. &\\(LightGBM)& It is based on decision trees. &\cite{lightgbm}\\
        \hline
        CatBoost & CatBoost is a popular open-source algorithm for gradient boosting&\\&  on decision trees. It includes an implementation of  &\\&ordered boosting, and an algorithm for &\\& processing categorical features in data. &\cite{catboost}\\
        \hline
        XGBoost & XGBoost is another Gradient Boosting algorithm,&\\& which performs tree boosting in parallel.&\cite{xgboost}\\
        \hline
        Neural Network & A neural network mimics the way the human brain operates by &See \cite{neural_network_review}\\
         (NN) &connecting a system of connected nodes (neurons) connected &for a recent\\& by edges. The inputs to neurons are real numbers, and the output &review\\&of each neuron is computed by some non-linear function & \\& of its inputs. The separate neurons are combined &\\& with different weights.& \\
        \hline
        Recurrent Neural  & This is a NN where node edges/connections have a direction to them&\\Network  (RNN)& This means the RNN can display temporal dynamic behavior. & \cite{rnn}\\
        \hline
        Gated recurrent & A popular variant of RNN, GRU includes gating units which&\\ unit (GRU) & control the amount of information flow inside each recurrent unit&\\& (controlling when the hidden state in a unit should be updated). &\cite{gru}\\
        \hline
        Multilayer perceptrons &MLPs are NNs where each neuron in one layer &\\ (MLP)& is connected to all neurons in the next layer. & \cite{Rosenblatt1963}\\
        \hline
        Convolutional Neural  & CNNs are regularized versions of MLPs &\\ Network (CNN) & and are commonly used on images. &\cite{cnn}\\
        \hline
    \end{tabular}
    \caption{A brief description of the methods used and acronyms in \acro.   \label{tab:acronyms}}
    \end{center}
\end{table*}
A few groups described fitting for hundreds of features and then using feature importance ranking to reduce the feature space over which classification was performed. 

\textbf{Data augmentation --} Central to the \acro~challenge was data augmentation, the supplementation of training data with new light curves derived through small changes to the original training set, and other approaches to mitigate non-representativity of the training set and class imbalance in both the training and test sets. Several groups tried a range of approaches to augment the data and rectify the class imbalance, which are discussed in more detail in Section~\ref{sec:challenges}, where we illustrate the degeneracies between the \classNINENINE~objects and the rest of the classes). 

\textbf{The \classNINENINE~objects --}
One of the most popular methods of computing probabilities for the\classNINENINE~was to probe the LB. All of the top models that provided information on their handling of \classNINENINE~objects used LB probing. One approach (`Weighted Average') produced a \classNINENINE~probability formed from a weighted combination of the supernovae class probabilities, and then determined the best weights by optimizing over their score or position on the LB.  The `Tuned Combination' from the MTMN method (see Section~\ref{sec:MTMN}) used a power law combination of supernovae probabilities and tuned the power and multiplicative factor using the LB. They also had two different functions for Galactic and extragalactic objects. Ahmet Erdem found which classes that \classNINENINE~objects were \emph{not} like (TDE, KN, AGN, SNIa, SNIa-91bg) and applied a constant multiplication factor found through LB probing depending on which class had the highest probability outside of those classes. Four of the top participants used a method that was shown on a kernel on Kaggle that involved scaling all the class probabilities with weights based on the LB probing from another user (`Probability Scaling').

\textbf{Dependence on simulated cadence --}
While the \acro~data were simulated for one proposed observing strategy for the \lsst, we note that the features themselves will in general be strongly dependent on the particular cadence or observational strategy of a survey. We leave a full investigation of classification performance for different observing cadences to future work. In the subsections below we summarize in more detail the winning solutions. 

\subsection{\topplace}

\begin{figure*}
\begin{center}
\begin{tabular}{l}
\includegraphics[width=0.75\textwidth]{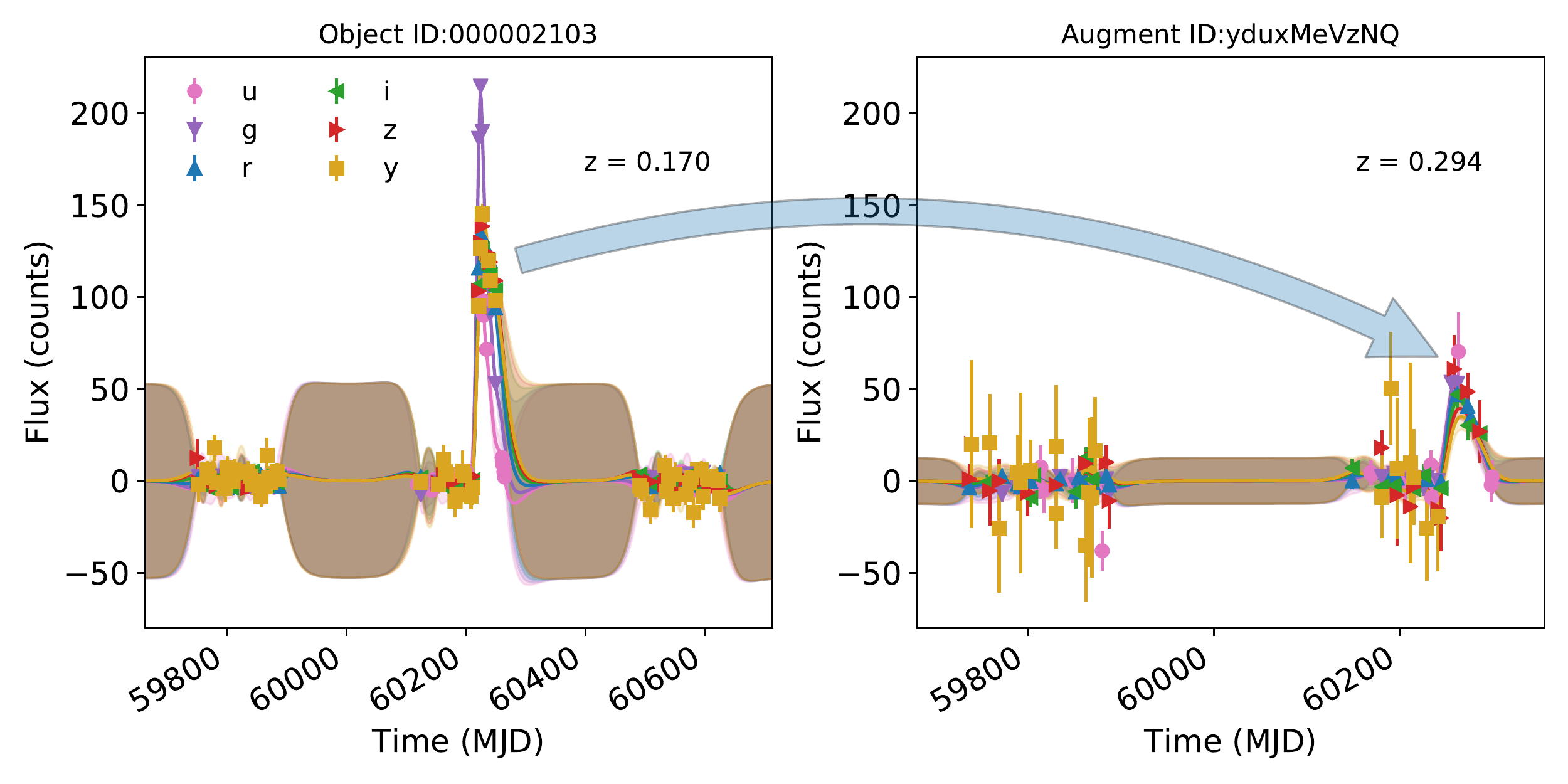}\\
\includegraphics[width=0.75\textwidth]{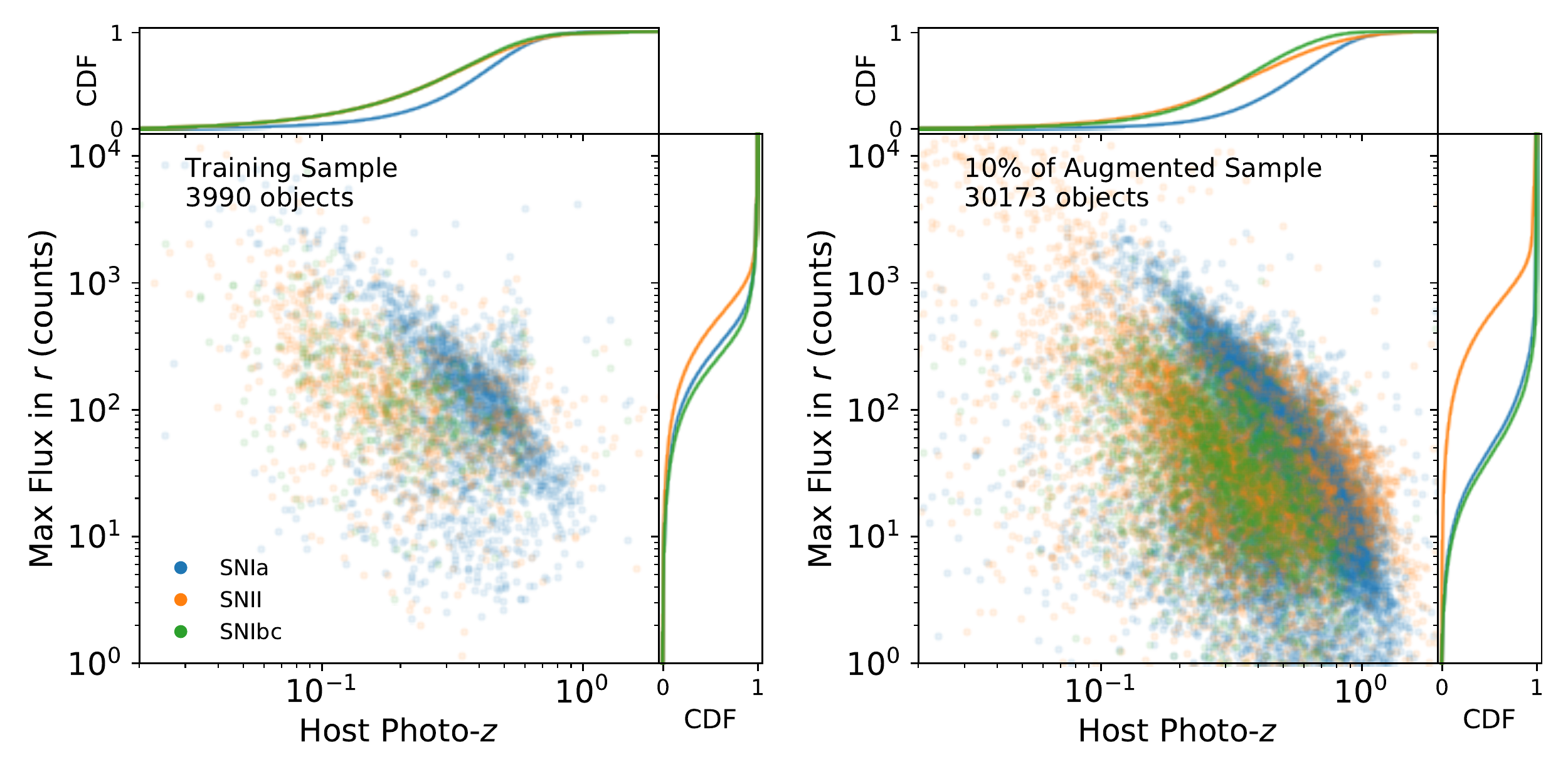}
\end{tabular} 
\caption{Data augmentation as part of the KB method. The light curves have been interpolated using GP interpolation, shown as shaded regions with largest spread where there is no data.  The top panel shows the augmentation in light curve space, where points are removed, the object is moved to higher redshift and the errors on the light curve points are increased. The bottom panel illustrates the augmentation in feature space for a subset of the \acro~models. The augmented sample not only has a larger spread in the maximum $r-$band flux, one of the features used by the {\texttt{avocado}} classifier but extends each class to higher redshift and helps increase the sample. \label{fig:augment}}
\end{center}
\end{figure*}

The overall winner of \acro~was Kyle Boone, then a graduate student at the University of California at Berkeley. His method using GP light curve fitting, data augmentation and a gradient boost \citep{gradient} classification strategy is described in detail in \cite{boone:2019}, which presented a slightly improved version of his submission to the \acro~competition. For the discussion here, we focus on the version of \texttt{avocado} that was submitted to \acro.
Boone was the only astronomer in the top three finishers, and his astronomical domain knowledge played a role in the choices he made for feature selection. For example, Boone extracted the color of the object at maximum light and computed the time to decline to 20\% of maximum light of the transient, which resulted in a greater accuracy of classification of SNIa-like objects in \texttt{avocado} than other challenge entries.

Pre-maximum observations are a useful indicator of the type of a supernova. As a corollary, the lack of pre-maximum observations can serve as a useful flag of how discerning the data will be in separating out different classes, and hence are a measure of the accuracy in classification of the simulations. \texttt{avocado} includes a flag for `incomplete' simulated light curves, which can help remove spurious classifications that might degrade the overall metric score. 

Data augmentation also features strongly in \topplace. The non-representativity of \acro~led to a pristine but small training data set and a large (3.5M-object sized) test data set, with much larger measurement uncertainties. Boone focused on augmenting the training data by shifting the data in time (without changing the shape of any light curve features), removing or `dropping' observations at random time positions. Boone also increased the scatter in the data to make it more representative of the test data quality, as shown in Figure~\ref{fig:augment}. In order to supplement the training data, Boone generated high-$z$ light curves from low-$z$ analogues by degrading the signal-to-noise ratio as a function of distance. This data augmentation greatly improved the accuracy of his \texttt{avocado}. The change of light curve colors with redshifting of the SED was not included in \texttt{avocado}.

\subsection{\secondplace}

The second place winners of \acro~were from a pair of competitors (Mike and Silogram) who joined forces to produce an ensemble solution of seven Recurrent Neural Network classifiers \citep[RNN,][]{rnn} and two Light Gradient Boosting Machine \citep[LightGBM,][]{lightgbm} models. 
Rather than focus on selecting one classifier, the ensemble provided diversity across model space, to avoid over fitting the training data. 

The RNN model was a one layer bi-directional Gated Recurrent Unit \citep[GRU,][]{gru}. 
The 7 different RNN models used between 80 and 160 units. 
The input data for the RNN was the flux and its associated uncertainty, the intervals between the measurements, the wavelength and passband, which were included with `one-hot' encoding, where combinations of the data are only those with a single value at a time  \citep[see][for details on this method]{onehot}. 
The models reduced the dimensionality by downsampling the layers and then combined the max pooled layer with light curve metadata, specifically the host galaxy photo-$z$ and associated error, the Milky Way extinction, and flags for the observational area/field.

As with the other solutions, the issues of data augmentation and feature selection were perhaps the most pressing for \secondplace. This method was supplemented to include augmentation near the closing date for submissions to the competition and so did not have the opportunity to retrain on a comprehensively augmented data set. The \secondplace~method dropped 30\% of the measurements and adjusted the redshifts and the flux of the training data. 

In the LightGBM models, adjusting the flux values according to the photometrically-derived redshift (photo-$z$) value helped `normalize' the flux, effectively correcting for cosmological redshift of the object, which stretches spectrum of the object and dims its brightness.  While the photo-$z$ was provided in \acro, the spectroscopic redshift (spec-$z$) is the more useful quantity as it is free of modeling errors. For both the RNN and the LightGBM models in this submission, the authors were able to build a separate model to predict the spec-$z$ of the host galaxy from the other properties of the object. This approach is useful particularly to flag potential outliers in photo-$z$ space (incorrectly predicted photo-$z$ values)\footnote{This is described in a note on the Kaggle discussion board \href{https://www.kaggle.com/c/PLAsTiCC-2018/discussion/75059\#latest-462457}{\textbf{online}}.}. This photo-$z$ modeling will be one of the key systematic issues that the Rubin Observatory will face, and any mitigation of this issue by classifiers is very valuable.
The \secondplace~authors stated that they converted all time and wavelength-related features to redshift-independent versions before training the RNN.

\subsection{\thirdplace}
\label{sec:MTMN}

The third place solution also went to a combination of teams, `Major Tom' and `mamas', themselves made of individuals with different implementations that were combined. In addition, `nyamp' provided feature engineering expertise to the combined group, hereafter known as the MTMN method.
The final solution was a 1D Convolutional Neural Network (CNN) with 256 8,5,3 convolution neurons followed by global max pooling as described above, with stride of unity. 
The inputs to the CNN in this case was a series of 128-long vectors over different `channels'. 
Six of these channels were constructed out of linear interpolated flux values (as a function of time) in the 6 bands, while another six were made of linearly interpolated vectors of flux multiplied by time, which forms a crude integral in the band. 
The final 6 channels in the set described the distance of the current point (in time) from the true input, effectively providing a vector describing the interpolation error.  

Three different CNNs were constructed based on different metadata features sent to the multilayer perceptron (MLP) which was the basis for the backpropagation: 
a minimal CNN based on using only the galaxy photo-$z$, distance modulus, MJD and flux error as metadata; a minimal CNN supplemented by the 16 best features obtained from features extracted in a separate CatBoost model framework; a minimal model supplemented by features obtained through template fitting from the Supernova Cosmology package \citep[\emph{sncosmo}, ][]{sncosmo}. 
The models are described in more detail on the Kaggle discussion boards\footnote{See e.g. the online links for the \href{https://www.kaggle.com/yuval6967/3rd-place-cnn}{\textbf{3rd place CNN}}, \href{https://www.kaggle.com/c/PLAsTiCC-2018/discussion/75131}{\textbf{3rd place CatBoost}}, \href{https://www.kaggle.com/c/PLAsTiCC-2018/discussion/75222}{\textbf{3rd place sncosmo}} entries.}. 

\begin{figure}
\begin{center}
\includegraphics[width=0.45\textwidth]{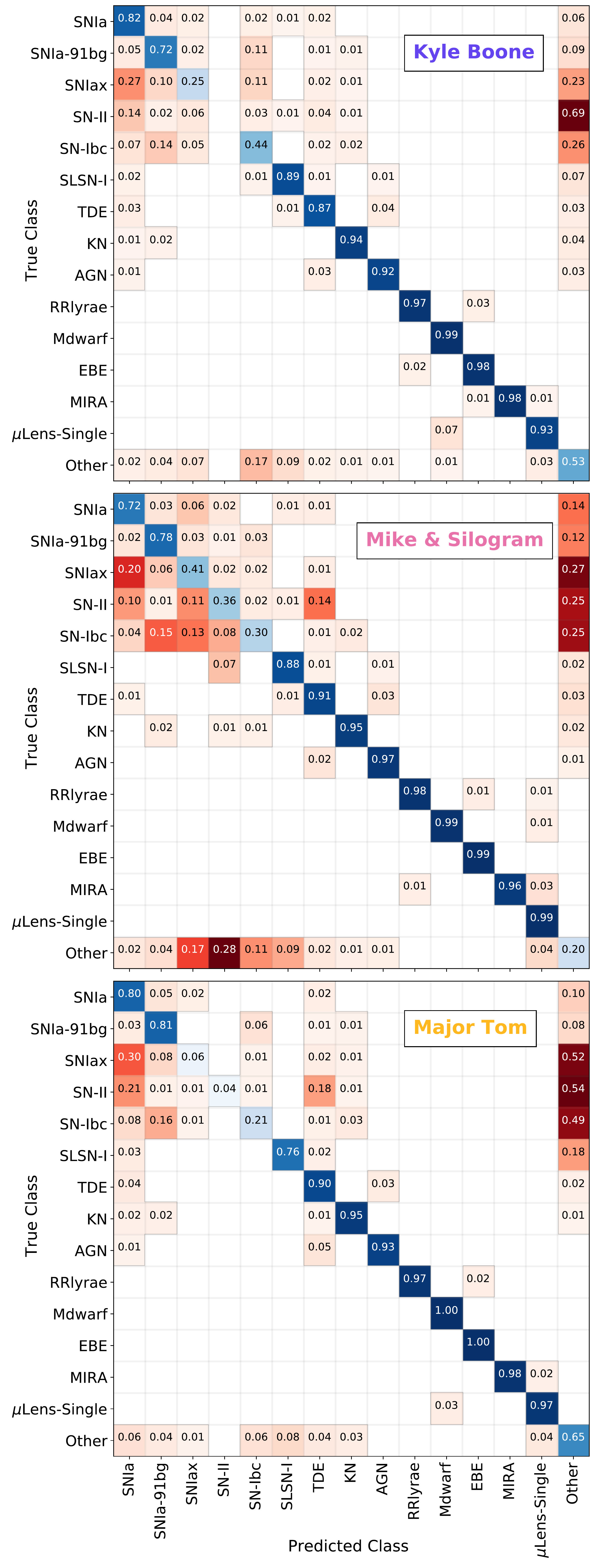}
\caption{Pseudo-confusion matrices for the top three entries to \acro.
The metric score for these entries was 0.68503 (\topplace), 0.69933 (\secondplace), 0.70016 (\thirdplace). The top entries are described fully in Tables~\ref{tab:methods} and \ref{tab:algorithms}.
\label{fig:confusion}}
\end{center}
\end{figure}

\begin{figure*}
\begin{center}
\includegraphics[width=0.8\textwidth]{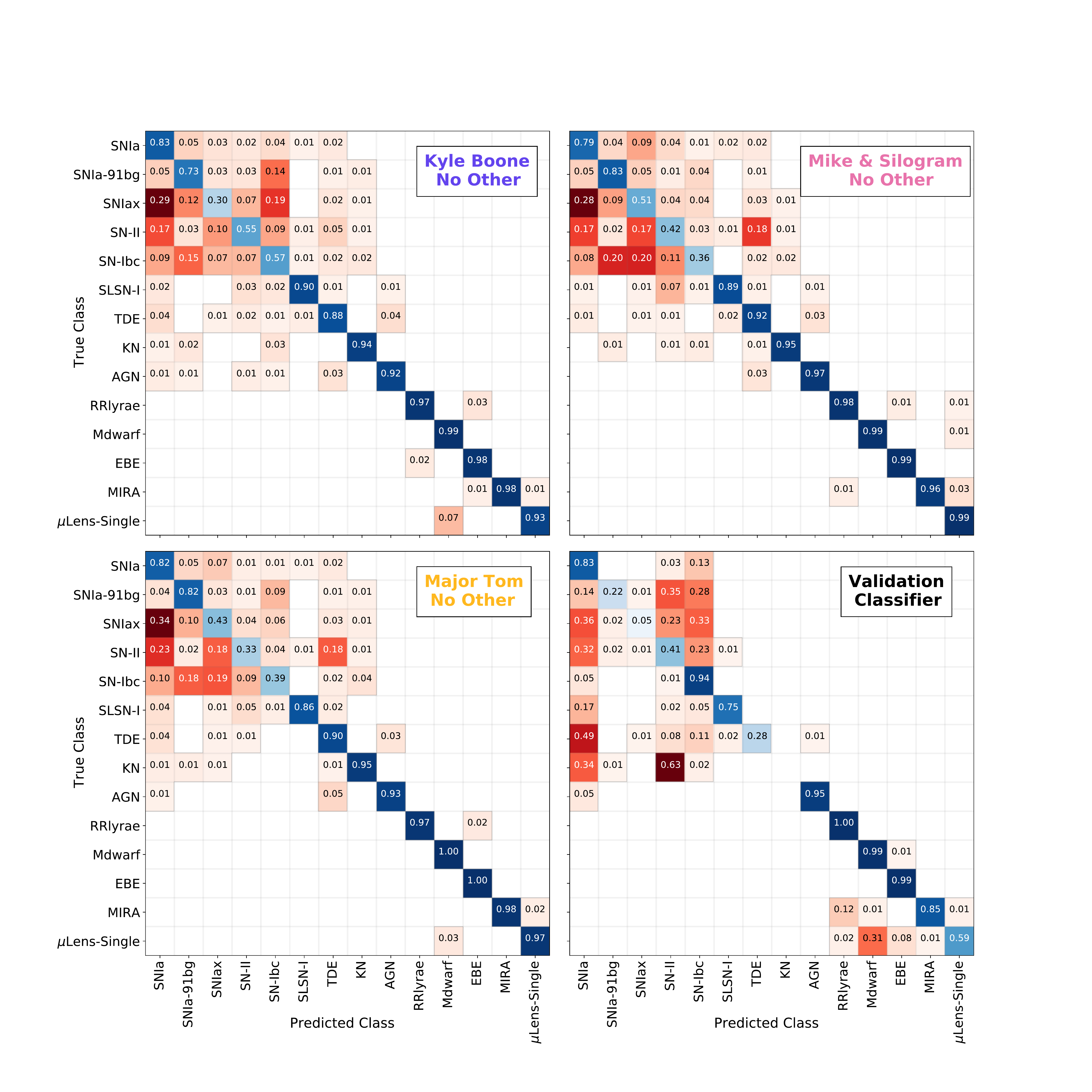}
\caption{Similar to Figure~\ref{fig:confusion}, but in this case the probabilities assigned to \classNINENINE~objects in the test set are omitted and the remaining probabilities are renormalized in the test data set. 
This illustrates more clearly the degeneracies between the known and the \classNINENINE~objects. The large `subsuming' classification systematic visible in Figure~\ref{fig:confusion} that arises since many classifiers provided a probability for \classNINENINE~objects based on a weighted sum of probabilities from the known objects is no longer present (consider the SNII-\classNINENINE~subsuming systematic in the \topplace~entry of Figure~\ref{fig:confusion}). We also show the results from the `validation classifier' run on the WFD data (without \classNINENINE~objects). 
\label{fig:confusion_no99}}
\end{center}
\end{figure*}

\section{Comparison of top-ranked classifiers by additional metrics}
\label{sec:alternative_metrics}

The \acro~metric was designed to evaluate submissions of classification probability mass functions (PMFs) without favoring any particular science case. For example, cosmological classification efforts have traditionally focused on removing contamination of non-Ia objects to yield a `pure' SNIa photometric sample, but metrics that reward only the purity of this class would be inappropriate for other science goals, such as novelty detection.

The caveats to the nominally science-agnostic global metric were twofold.
The classification of rare and novel classes was encouraged by increasing the relative weight of the contribution of these classes to an overall metric that was otherwise evenly weighted across classes. This division into classes did not encode class hierarchy; had all supernova sub-classes been combined into a single class, or had galactic and extragalactic classes been evaluated separately, it is possible that a metric would have favored different classifiers.

While the Kaggle metric was required to be a scalar so that a cash prize could be awarded to an overall winner, the resulting classifications in the form of submitted PMF catalogs are a much richer data set. 
In this section, we compare the top-ranked classifiers by more nuanced and higher-dimensional metrics of estimated classification PMFs without investigating metrics tuned to any specific science goals. We leave such an exploration for future work.

\subsection{Pseudo-Confusion Matrices}

The confusion matrix is a classic metric of deterministic classifications \citep[see e.g.][for one such astronomical example]{snmachine}. 
To put the \acro~results into context for those familiar with the confusion matrix, we present an adaptation thereof to probabilistic classifications.
For each classifier, we assign each light curve a deterministic classification corresponding to the mode (maximum) of its PMF.
To remove the visual domination of class imbalance, we normalize the rows (corresponding to predicted classifications for a given true class) of a standard confusion matrix, whose cells contain counts rather than proportions relative to the number of objects in the true classes.

The pseudo-confusion matrices for the solutions of the top three classifiers are shown in Figure~\ref{fig:confusion}.
The matrices show a distinct difference in the performance of the classifiers when distinguishing between different types of supernovae, which are clustered in the top left-hand corner of the diagram. 
In addition, the degeneracy between the classification of true \classNINENINE~objects (right column) and Type II supernovae (fourth row of the matrix) is shown in Figure~\ref{fig:confusion}. The degeneracy arises because three of the four novel objects included in the~\classNINENINE~had supernova-like light curves and were difficult to disentangle from the supernovae. In addition, some entrants formed a composite classification probability for this class from a weighted average combination of probabilities of the other classes (notably the SNIbcs and the SNIas), leading to correlation between the classification probabilities of specific objects and the \classNINENINE.

We show the confusion matrices considered \textit{without} the \classNINENINE~objects in Figure~\ref{fig:confusion_no99}.
The classifiers were not rerun without the \classNINENINE~objects. We removed the \classNINENINE~objects from the submissions and renormalized the probabilities. None of the classifiers shown here attempted to classify \classNINENINE~objects through their machine learning algorithms but instead combined the probabilities from other classes. The validation classifier was also run without \classNINENINE~objects such that it is a better comparison to these confusion matrices than the ones in Figure~\ref{fig:confusion}.

\begin{figure*}[htbp!]
\begin{center}
\begin{tabular}{ll}
\includegraphics[scale=0.4]{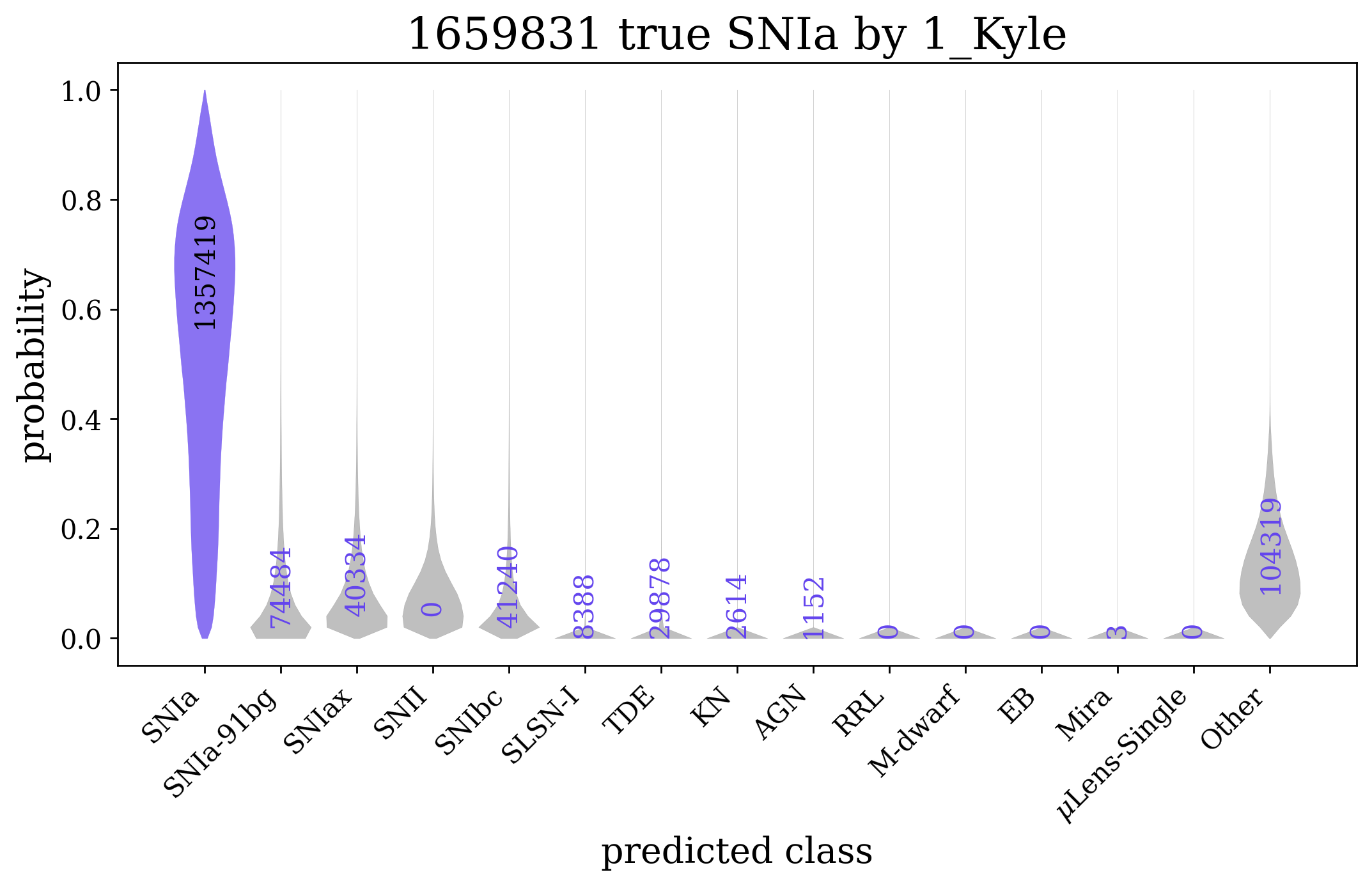} & \includegraphics[scale=0.4]{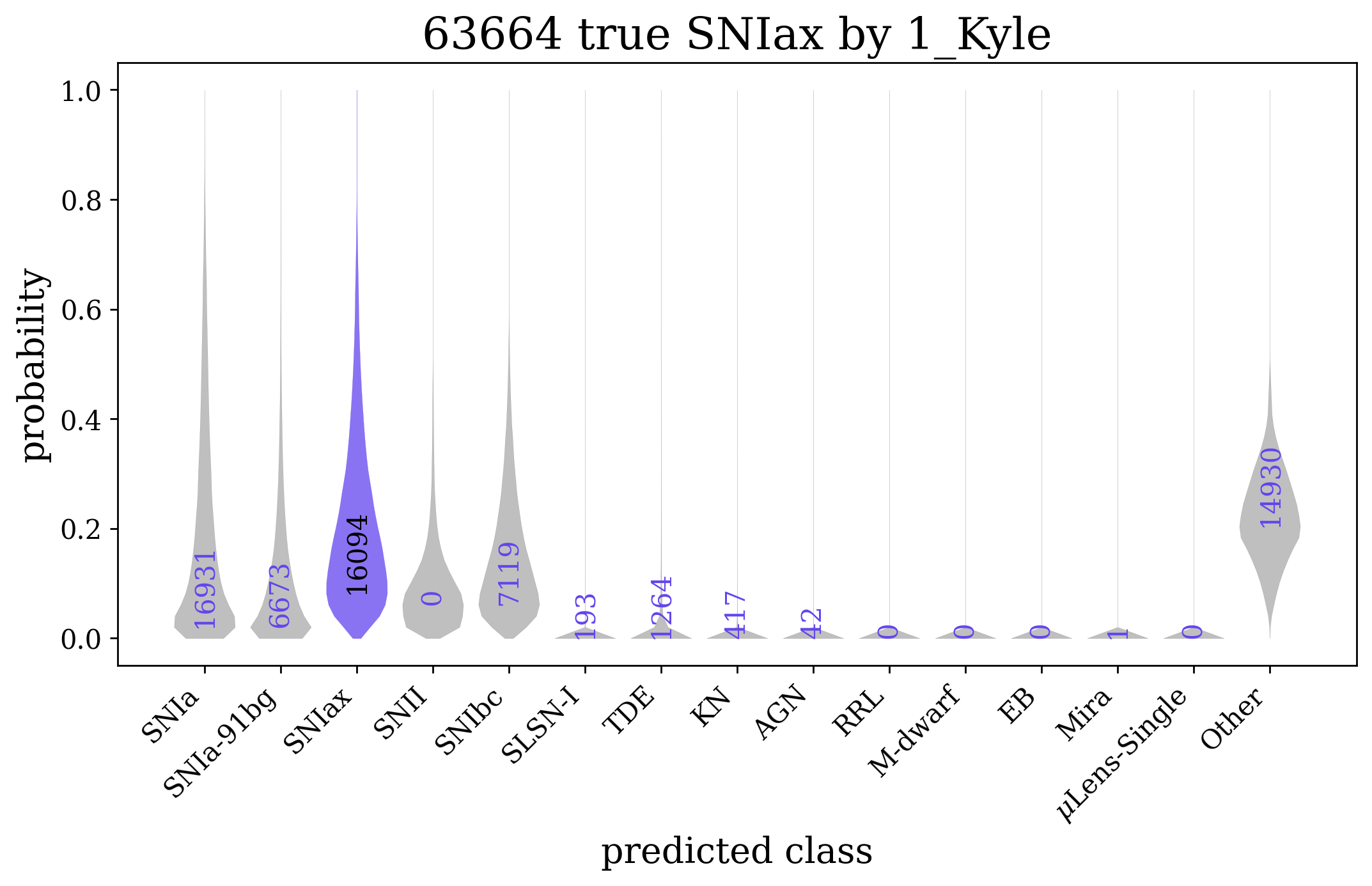}\\
\includegraphics[scale=0.4]{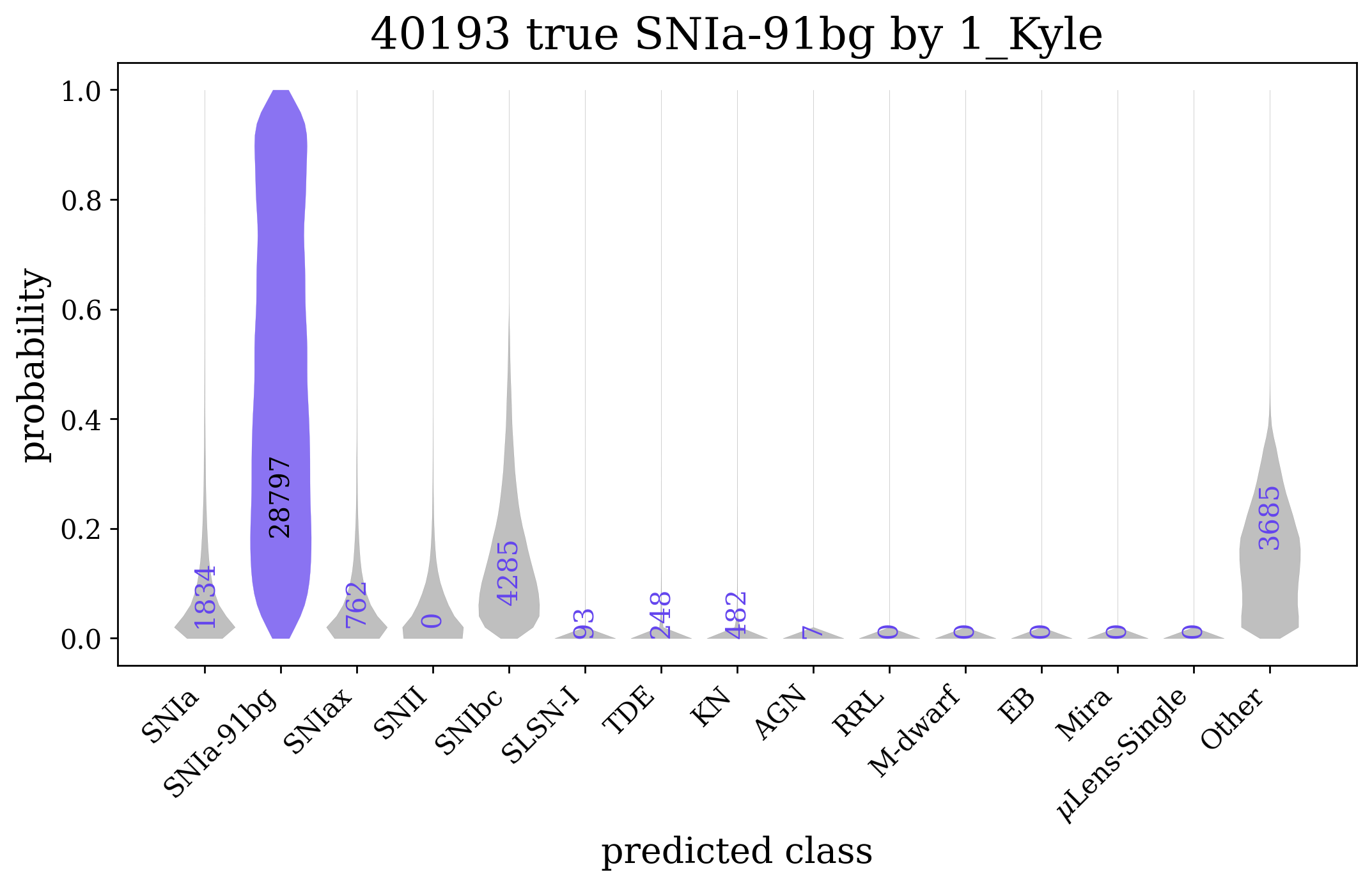} & \includegraphics[scale=0.4]{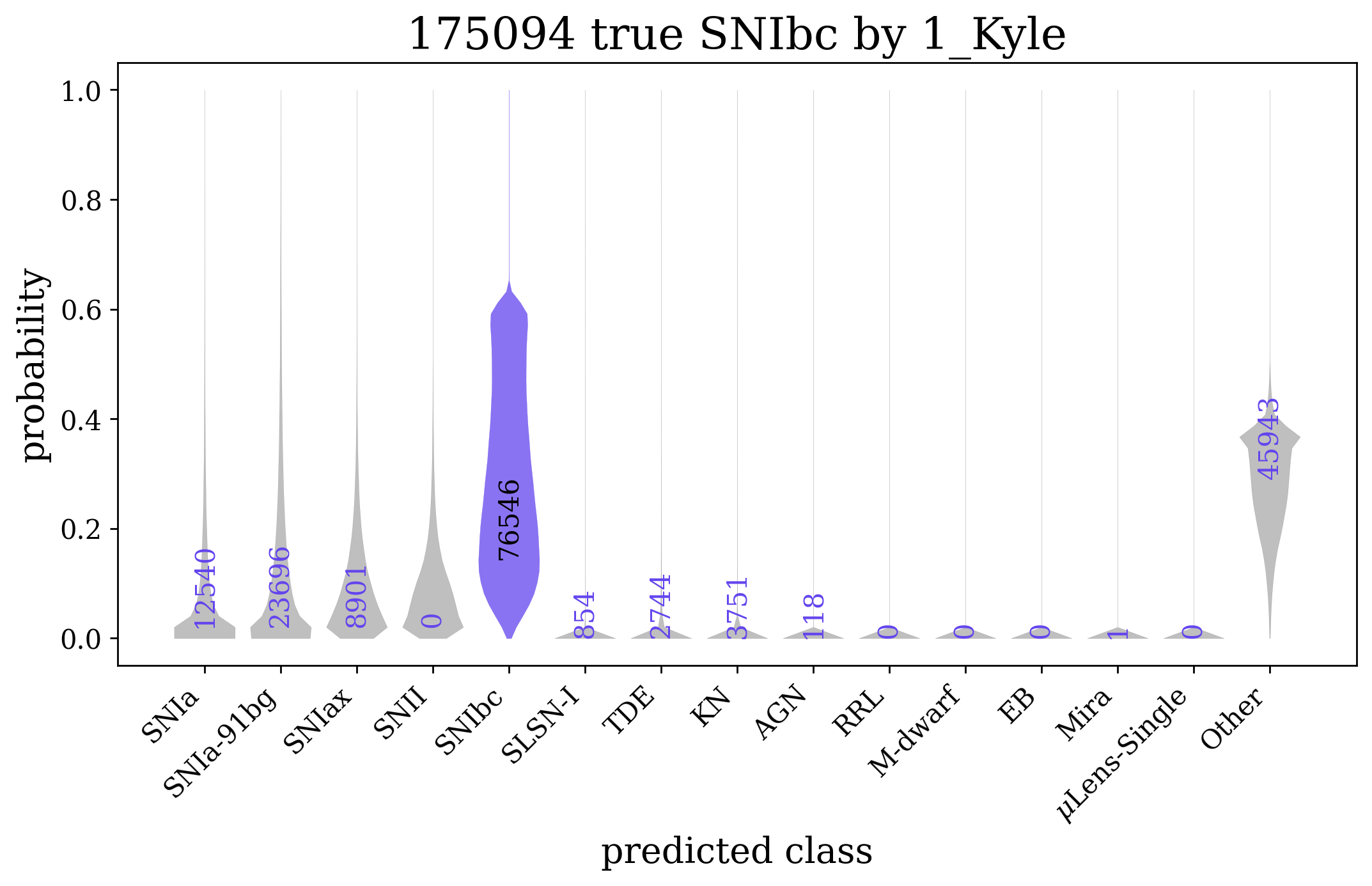}
\end{tabular} 
\caption{The estimated posterior probabilities (violins) for each predicted class (columns) among a given true class (panels) for the \topplace~method, where the assigned class corresponding to the true class is highlighted in color. 
\textbf{Top row:} The estimated probabilities for well-classified SNIa and those of the less well-classified SNIax. 
The assigned probability of being the true class for SNIa has a long tail but a high mode of $\sim0.8$, whereas SNIax are degenerate with the \classNINENINE~type, whose mode is $\sim0.2$.
\textbf{Bottom row:} The classification probabilities for SNIa-91bg and SNIbc (core-collapse SN).  
The estimated classification probabilities for true SNIa-91bg objects reflects uncertainty; 
while roughly two thirds of the true SNIa-91bg objects have a mode corresponding to the predicted SNIa-91bg class, their classification probabilities span the entire range of probability. 
The uncertainty of estimated probabilities for SNIbc objects manifests differently; 
the probability for predicted class SNIbc has a less pronounced tail, but there is considerable probability mass assigned to predicted \classNINENINE. 
This is not surprising given that the \topplace~classifier determined the \classNINENINE~classification probability from a combination of the probabilities of the SNIa, SNIbc, and SNII. 
\label{fig:violin}}
\end{center}
\end{figure*}

\begin{figure}[htbp!]
\begin{center}
\includegraphics[scale=0.4]{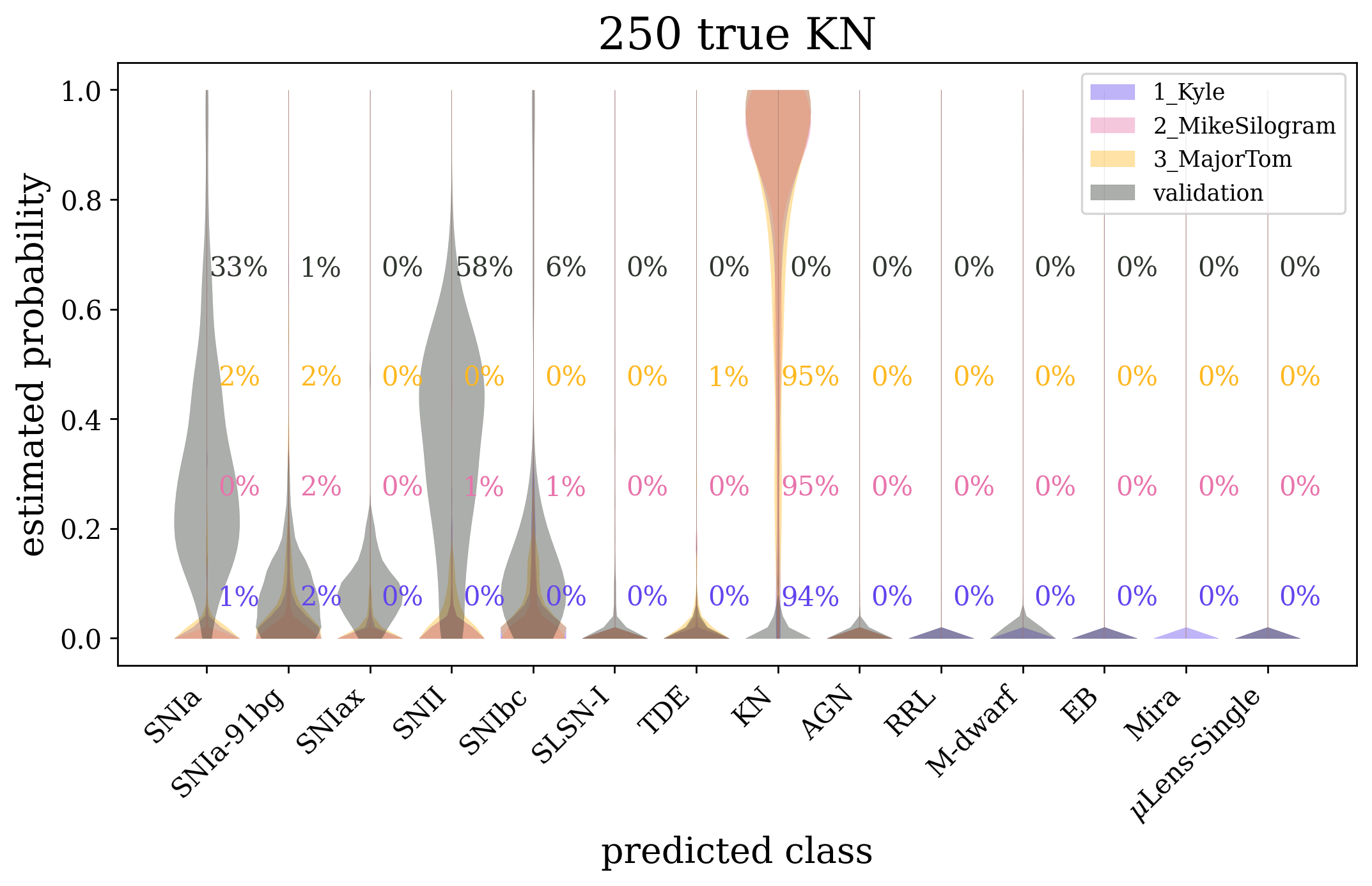}
\caption{Performance of the `validation' classifier compared to the top three entries to \acro. The na\"ive classifier does not correctly classify the rare KN objects, while a classifier more optimized to find rare objects through data augmentation is a very accurate classifier of these bright transients.\label{fig:violin_validation}} \end{center}
\end{figure}

\begin{figure}[htbp!]
\begin{center}
\includegraphics[scale=0.4]{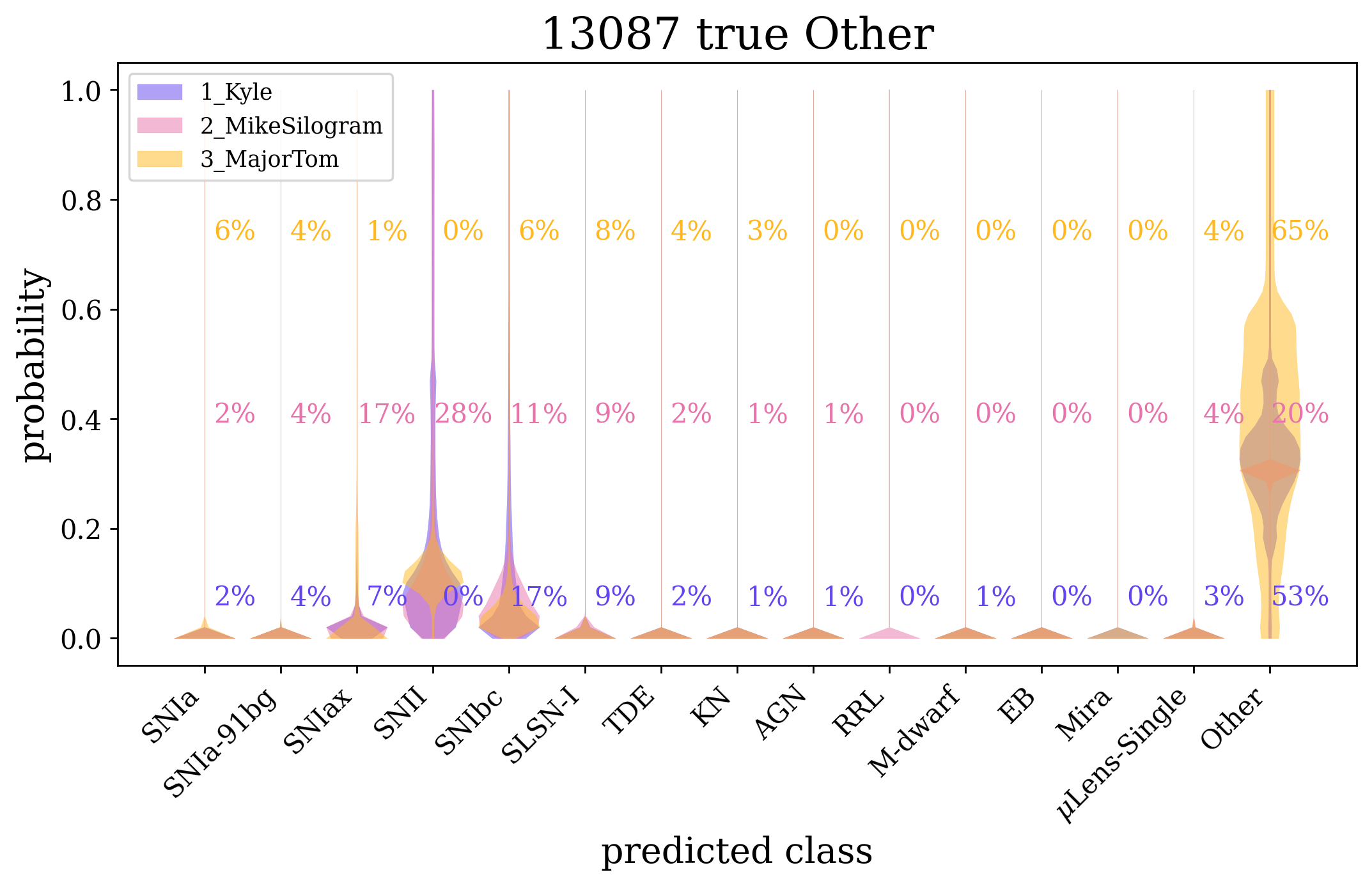}
\caption{Performance of the top three classifiers for the \classNINENINE. The predicted probability distributions for the 13087 true type \classNINENINE~objects. The numbers in different colors reflect how many objects were incorrectly classified as a given type by each of the three classifiers, and the shaded regions show the predicted probability distributions for each class. Note how the \classNINENINE~is most often confused with SNII and SNIbc. This occurred as a result of many of the classifiers generating their \classNINENINE~probability as a combination of probabilities from the other classes.   
\label{fig:violinall}}
\end{center}
\end{figure}

\begin{figure*}[htbp!]
\begin{center}
\begin{tabular}{l}
\includegraphics[width=0.8\textwidth]{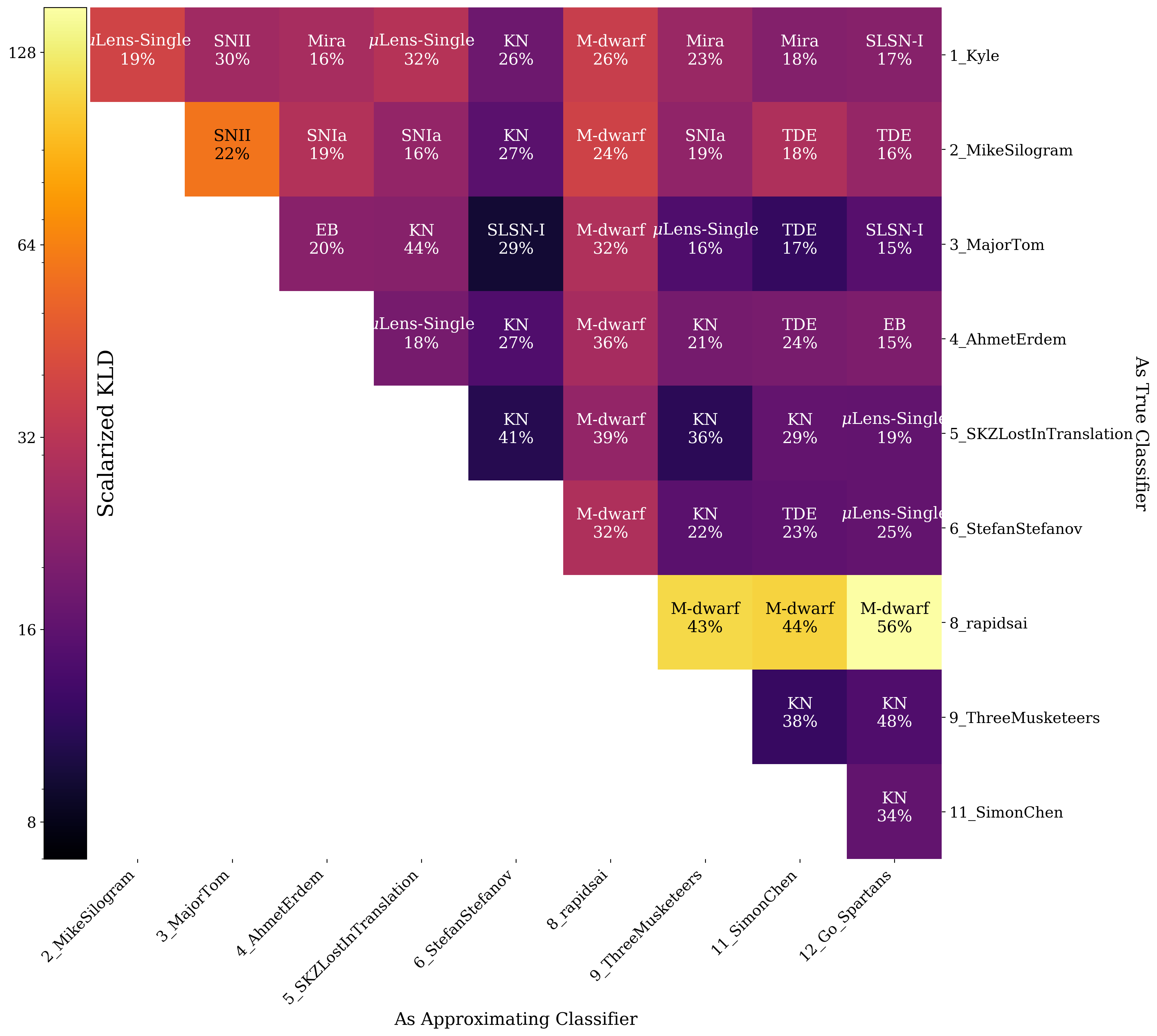}
\end{tabular} 
\caption{The disagreement matrix for the top classifiers. 
Colors indicate the disagreement metric between each classifier ($x$-axis) and all classifiers ranked better than it ($y$-axis), where darker cells correspond to lower Kullback-Leibler Divergence from the worse-ranked classifiers' probability densities of assigned posterior classification probabilities to those of the better-ranked classifier. The higher- and lower-ranked classifier pairs' cells are labeled with the class that was most different between the two classifiers and percentage of the \textit{total disagreement between the two classifiers} for which that class is responsible. The methods used by each classifier are summarized in Table~\ref{tab:methods}.  
\label{fig:kld}}
\end{center}
\end{figure*}

\subsection{Distributions of assigned classification probabilities}

\acro~required submissions of posterior PMF vectors comprising a much richer set of resultant data than solely deterministic metrics such as those related to the numbers of object classified as true/false positives/negatives.
The violin plots of Figure~\ref{fig:violin} depict the probability density functions (PDFs) of estimated PMF values across all predicted classes for each true class's light curves, illustrating more than just a summary statistic of the PDF of estimated posterior PMF values, such as the mode or mean. 
Though the per-light curve covariance is lost, such a visualization still conveys some of the degeneracies between different true and predicted classes, like an expanded form of each row in a confusion matrix. 

For a light curve of a given true class, one would hope that a classifier would assign it a high probability for being of that class and low probabilities for being all other classes.
Two panels of Figure~\ref{fig:violin} show that the winning classifier exemplifies this desirable behavior for true SNIa and SNIbc by assigning high probabilities for their true class and low probabilities for all other classes. 
The degeneracies between classes can be seen as PDFs of the estimated posterior PMFs where there is substantial density for a predicted class that is not the true class (i.e. the peak of the PDF is skewed away from zero for a class other than the true class). 
The remaining two panels of Figure~\ref{fig:violin} provide illustrations of this undesirable effect: objects with true SNIax and SNIa-91bg type are assigned comparable probabilities for several classes including the true class. Although the centering of the PDF density around zero PMF for some mismatched classes indicates that the winning classifier did indeed easily rule out those incorrect classes, there is also low PDF density at high PMF for the true class (consider that the PDF drops off sharply for probability $>0.15$ for the SNIax in the top right panel of Figure~\ref{fig:violin}), meaning that the classifier was unable to distinguish between those SN sub-types.

Figure~\ref{fig:violin_validation} isolates the true KN to compare the performance of the top three classifiers to that of the na\"ive `validation' classifier, whose inclusion is facilitated by including only light curves evaluated by the validation classifier that also appeared in the \acro~data set and omitting the probabilities assigned to the \classNINENINE~prior to renormalizing the probabilities assigned to the remaining classes.
All of the top three classifiers were able to classify the true KN objects, as they assigned high PMF values for the predicted KN class and low PMF values for all other classes.
The validation classifier, on the other hand, assigns nonzero PMF values over all the SN sub-types, effectively mis-classifying KN as SN without being able to distinguish the SN sub-type.

Figure~\ref{fig:violinall} shows the performance of the top three classifiers on the \classNINENINE. 
All three top classifiers successfully isolated the \classNINENINE~light curves from most of the named classes and for the most part successfully assigned a higher posterior PMF value for the correct class.
However, all three top classifiers also had high PDF density at nonzero PMF values for SNII and SNIbc, meaning that they frequently allocated nontrivial probability to the possibility that~\classNINENINE~light curves were SNII and SNIbc.

\subsubsection{Classifier disagreement}
\label{sec:scalarized_kld}

When comparing classifiers, it is useful to establish a notion of general disagreement between the estimated posterior PMFs across classifiers and classes on the basis of information content.
If we had access to the true PMF values corresponding to the \acro~ data set, we could measure the loss of information due to using an estimated PMF $g(x)$ rather than the true PMF $f(x)$ using the discrete Kullback-Leibler divergence (dKLD)
\begin{equation}
    \label{eqn:dkld}
    \mathrm{dKLD}(f, g) = \sum_x f(x) \left( \log[f(x)] - \log[g(x)] \right)  ,
\end{equation}
where $x$ takes categorical values corresponding to each class, and the sum is over the set of $x$ values for which $f(x) > 0$.
To summarize these per-light curve values, one can calculate the continuous Kullback-Leibler divergence (cKLD) of the PDFs $g(x)$, $f(x)$ of approximating PMFs relative to those of the true PMFs, respectively, 
\begin{equation}
    \label{eqn:ckld}
    \mathrm{cKLD}(f, g) = \int f(x) \left( \log[f(x)] - \log[g(x)] \right) dx ,
\end{equation}
where $x$ is now the set of real numbers between $0$ and $1$.
In either case, the KLD is measured in nats, the unit of base-$e$ information.
The appendix of \citet{Malz_2018} is a good primer on the KLD including a demonstration to build intuition for the behavior of the KLD in limiting cases.
The true PMF is approximated from the Kaggle metric as follows. If we were to assume that the Kaggle ranking metric correctly orders classifiers by how close their submitted PMFs are to the (unknown) true PMFs, then this ordering is sufficient to define \textit{a directionality} for evaluating the KLD.\footnote{The validity of this assumption is equivalent to $\rm{KLD}[i+1; i-1] = \rm{KLD}[i+1; i] + \rm{KLD}[i; i-1]$, which is formally untrue for this data set, but the assumption is useful in this context.}

To visualize the disagreements between classifiers, we convert the KLD values computed per true class per each predicted class and per light curve within a true-class. 
The cKLD of Equation~\ref{eqn:ckld} corresponds to the divergence for one pair of true and predicted class.
The per-true-class sum of the cKLD over the PDFs of per-predicted-class PMFs is interpretable as a measure of the information about that true class. 
A final scalar value may be made by taking the weighted sum of these per-true-class cKLD terms using the per-true-class weights of the original Kaggle metric (as defined in Table~\ref{tab:weights}). 
This integrated KLD may then be interpreted as the value of the information loss due to using the approximating classifier rather than the one that is closer to the truth.
The results of a summation of the pairwise KLD between the per-true class, per-predicted class PDFs of PMF values of each classifier as an approximation to those of all classifiers ranked better is presented in Figure~\ref{fig:kld}.

The scalar KLD values of the pairs span three orders of base-$e$ magnitude;
however, if we exclude the 8th ranked classifier, then the pairwise information losses in the PDFs of PMFs only span two base-$e$ orders of magnitude.

The reason for this large range in KLD values is that the affected classifier  (`8\_rapidsai') assigned nearly deterministic classifications to the M-dwarf class, meaning the PMF values tended to be approximately 1 for one predicted class and approximately 0 for all other predicted classes.
This choice induced long tails in the per-assigned class PDFs of true M-dwarf PMF values, to which the KLD is known to be sensitive, particularly when the PDF taken as closer to truth is affected.

Figure~\ref{fig:kld} illustrates a few additional trends. The top two classifiers disagree \textit{more strongly} with all other classifiers than the other classifiers do with one another.
The class for which the discrepancy is greatest varies quite a bit among classifier pairs, but aside from the outlier classifier discussed above, the KN and Mira objects appear most discrepant often, indicating that the other ranked classifiers' strategies resulted in very different classification performance on those classes in particular.

\subsection{Precision-Recall Plots}
\label{sec:pr}

\begin{figure*}[htbp!]
\begin{center}
\begin{tabular}{l}
\includegraphics[width=0.75\textwidth]{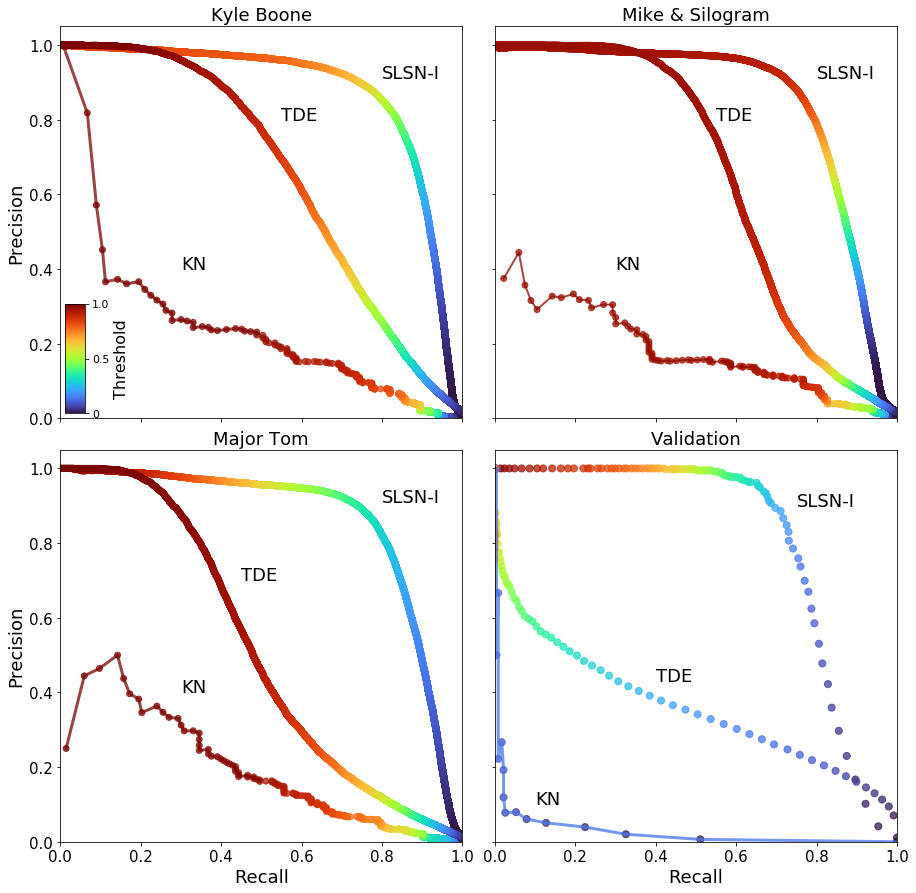}\\
\end{tabular} 
\caption{The precision-recall plots for the top three classifiers and the `validation' classifier as a function of threshold (color) for three classes of interest: KN, TDE, and SLSN.
\label{fig:prc}}
\end{center}
\end{figure*}

Deterministic classifiers are often evaluated using functions of the true and false positive and negative rates. The true positive (TP) rate is the ratio of TP classifications to the sum of both the TP and false negative (FN) classifications $ N_\mathrm{TP}/(N_\mathrm{TP}+N_\mathrm{FN}$).
The false positive (FP) rate is the ratio of the FPs to the sum of both FP and true negative (TN) rate. 
One example of such a function is the receiver operating characteristic (ROC) curve, the TP rate as a function of the FP rate for a given classifier. 

Another is the precision-recall (also referred to as `completeness-purity') plot of the precision $P = N_\mathrm{TP}/(N_\mathrm{TP}+N_\mathrm{FP})$ as a function of the recall $R = N_\mathrm{TP}/(N_\mathrm{TP}+N_\mathrm{FN})$.
Recall is a performance measure of the TP compared to total \textit{actual positives}, whether TP or FN, whereas precision is a performance measure of TP predictions compared to the total positive predictions (whether true or false). We choose to use precision-recall plots over ROC curves because they are appropriate for imbalanced data sets \citep[see e.g.][]{saito/etal:2015}. 
The precision and recall are calculated for different values of the threshold of the classifier to map the performance of both as a function of threshold. We require higher classification \textit{confidence} with increasing threshold values. As the threshold increases, our precision \textit{increases} and recall (akin to sensitivity) \textit{decreases}. Good performance on these plots would be indicated by having high precision and high recall, which would indicate that the classifiers are returning a lot of labels that are accurate. 

A perfect classifier in these plots would have a precision of one for all recall. A bad classifier will output scores whose values are only slightly associated with the true outcome, and so will achieve high recall for low values of precision. 
Figure~\ref{fig:prc} shows examples of precision-recall plots for three different objects classified by the top three classifiers and the validation classifier. For the top three participants, SLSN are more well-classified than KN or TDE but are not perfectly classified. 

The classifiers also did approximately the same in precision and recall for TDEs.
Of the transient events shown in this example, the KN are the most easily misclassified.  This object has low precision, meaning that true positive classifications are a small fraction of total positive classifications. As the threshold value increases (needed to improve the precision performance), the recall (true positives relative to true positives and false negatives) drops relatively slowly as the number of false negatives is high for the KN class. The greatest contaminants to KN classification are SNIa and SNII, which is seen in both the \acro~simulations and in the Dark Energy Survey observations \citep{doctor/etal:2017,morgan/etal:2020}. The MTMN and MS methods have a poor performance on this class throughout the plot whereas the KB method has good early ``retrieval''. 

The SLSN have high values of recall for a range of thresholds, with $P > 0.9$ for all thresholds $> 0.7$, indicating it is a well-classified object. TDE require a high threshold to be correctly classified, but then plateau at constant $P$ for threshold $\gtrsim 0.9$ for all classifiers. For the  SLSN, the validation classifier performed significantly better than the top three classifiers, reaching $P\simeq 1$ for thresholds $\gtrsim 0.25.$

\subsubsection{Redshift evolution of metrics}

\begin{figure}[htbp!]
\begin{center}
\begin{tabular}{l}
\includegraphics[width=0.45\textwidth,trim={0.7cm 1.5cm 0.7cm 0},clip]{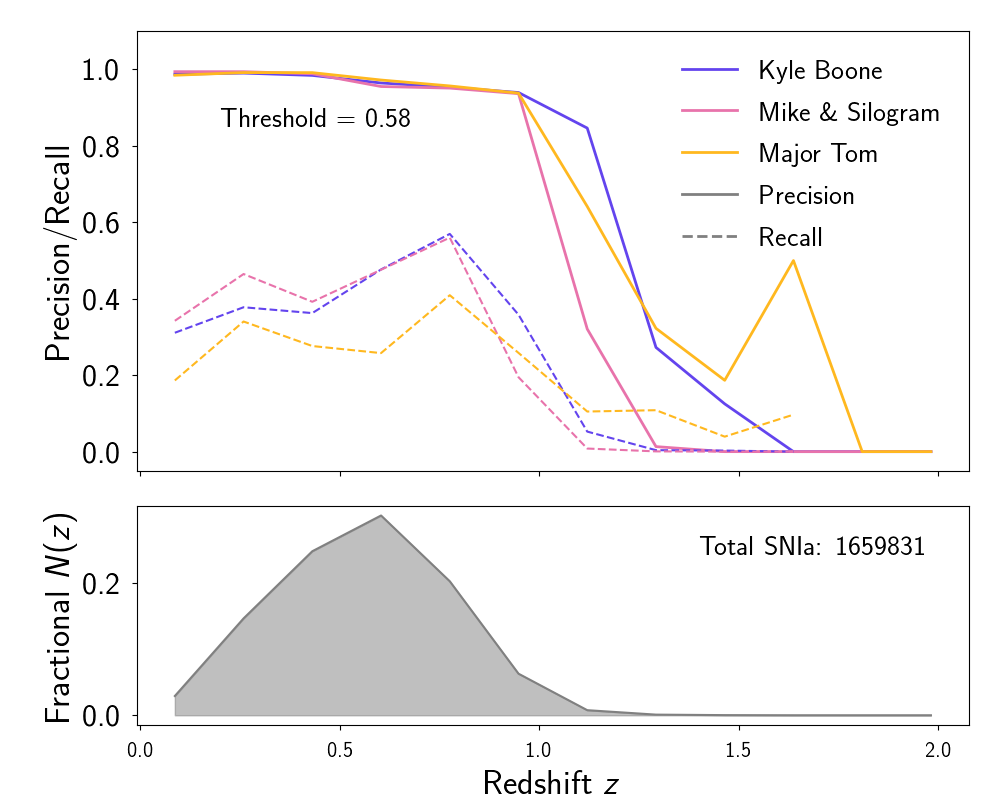}\\
\includegraphics[width=0.45\textwidth,trim={0.7cm 0.0cm 0.7cm 0.5cm},clip]{./figures/twopanel_pr_SNIa.png}\\
\end{tabular} 
\caption{The precision and recall as a function of redshift for the SNIa objects (top panels) and TDE (bottom panels). The average value of the threshold used across the classifiers is indicated in text on the figure. The shaded distributions in each panel show the fractional distribution of the numbers of the transients as a function of redshift. The total number of transients in the sample is also indicated. The evolution with redshift of the metrics depends both on the redshift distribution of the objects and the evolution of the light curves with changing redshift.\label{fig:accuracy} }
\end{center}
\end{figure}

While the global precision-recall plots for various classifiers and types are shown in Figure~\ref{fig:prc}, it is worth considering the evolution of the classifier performance with redshift. In Figure~\ref{fig:accuracy}, we show the precision and recall of the top three classifiers as a function of redshift for given threshold value of $\sim 0.6$ for two object types, namely SNeIa and TDEs. Also shown are the redshift distributions of the transients. The relative numbers of TDE peak at lower redshift (while there are still greater total numbers of SNIa than TDE at all redshifts), leading to a peak in recall. The precision of the classifier peaks later, with the corresponding drop in the recall. This behavior is also seen in Figure~\ref{fig:prc} which shows the precision-recall curves as a function of threshold value rather than redshift. The ability to classify objects well with increasing redshift is driven in part by the redshift distribution of the sources themselves, and by the quality of data augmentation of the training set. As noted in \cite{boone:2019}, prioritization of completeness of a given survey as a function of redshift will lead to rather different tuning of classifier performance.


\section{Discussion of challenges in the \acro~data}
\label{sec:challenges}

The \acro~data set was rich in complexity and highly non-representative. 

\textbf{Detection of novel objects --}
An additional class in the test set but not present in the training set, named the \classNINENINE, added the task of anomaly detection to the challenge. 
This unseen class proved challenging and led to confusion classifications that exhibit the `subsumed-to' behavior as described in \cite{malz/etal:2019_metric}, where a first class is consistently mistaken for a second, but the second class is not commonly mistaken for the first. 
In Figure~\ref{fig:confusion_no99} we show the confusion matrices for the same entries as in Figure~\ref{fig:confusion}. In this case while the \classNINENINE~objects were included in the test set, the classifiers did not return a classification for the objects (hence the predicted probabilities are set to zero for these objects). Hence while the true \classNINENINE~objects have predicted probabilities of the other classes by construction, the subsuming probability systematic where true objects are confused for \classNINENINE~objects is reduced. This can be most easily observed for the SNII objects in e.g. the \topplace~classifier. We do not investigate separately the confusion between the four sub-classes that made up the \classNINENINE, however future work and larger simulations could tease out the relative confusion of these interesting objects with well-known types.

This confusion between the other classes and the \classNINENINE~objects is illustrated in Figure~\ref{fig:violin} through violin plots of probability density of the posterior probability assigned to each predicted class, for a given true class. There are a few things worth noting in the violin plots. First, the distribution of probability extends to zero for some true objects, like the SNIa-91bg which is shown in the bottom left panel of Figure~\ref{fig:violin}, indicating that this true type is intrinsically uncertain. For the true type SNIbc, the violin plots are even more interesting. Here the violin plot for the \classNINENINE~has a peak at P(type) = 0.4 significantly away from zero. This arises from the fact that the assigned probability of the \classNINENINE~in Boone's solution was derived as a combination of probabilities from SNIa, SNIbc and SNII.  

\textbf{Feature selection -- } 
The participants in the \acro~challenge were unanimous that the biggest challenge they faced was the extraction of features from the large training and test sets of \acro. Feature extraction is often the limiting factor in these classification challenges. Given the existence and use of the \cite{bazin/etal:2012} light curve fitting model (which is a four-parameter model that tends to zero flux as $t\rightarrow \pm \infty)$ which is in many commonly-used astronomical codes, many groups started with fits to the model, and trained their classifiers on the fit parameters.
Consistently one of the top features was either the host galaxy photometric redshift or a proxy for the spectroscopic redshift modeled from the photometric redshift. Only two of the top ten \acro~entries \textit{did not} include an estimate of the redshift or the distance modulus. The maximum flux was also frequently used as a feature for classification. 
The \topplace~classifier included many signal-to-noise features whereas most other submissions traced the time-dependent movements of the light curves between the maximum and minimum flux.
Many people used some form of the ratios of flux in different filters (essentially a `color') as classification features.

GP regression proved a more model-independent approach, but added the complexity of ensuring the correct choice of kernel to do the regression. 

\textbf{Non-representativity of the \acro~test data -- }
\lsst~will map out the sky to a co-added magnitude depth of $27-28$ and a single-visit depth of $24$ in the various bands. This depth will yield a survey that has orders of magnitude more objects than current wide-field surveys. No current survey will be capable of providing distributions of classes as a function of redshift that will be representative for \lsst. Any classifications performed based on low-$z$ training sets will by definition be incomplete. To emphasize the importance of developing classifiers robust to this, the \acro~training set is non-representative by design. The training set of 7486 objects were chosen to represent a probable spectroscopic survey made from a combination of objects from the Dark Energy Survey (DES) and other low-redshift surveys like the Foundations survey \citep{foley/etal:2018_foundation}. The procedure for determining the spectroscopic training sample is outlined in Section 6.4 of \cite{kessler/etal:2019_PLAsTiCC}. The roughly 3 million objects in the test set not only had a different redshift distribution, but included the \classNINENINE~as a way to test robustness of classifiers to new objects and provide a sample for anomaly detection, as discussed above.

A small fraction of the 3 million test objects included spectroscopic `confirmation' and therefore had a small redshift error, while others had their redshifts assigned according to an example of the projected \lsst~ photometric redshift performance. A small bias to the photo-$z$ values is introduced as a result of the broad band data and 
training process. Many groups used the photo-$z$ and/or the flux information to determine a proxy for the more informative spectroscopic redshift, which was used to determine an effective distance for the objects. These are labeled as `Model spec-$z$' in Table~\ref{tab:methods}.
We leave it to future work to investigate the impact the photo-$z$ bias had on classification performance explicitly.  
Several groups performed data augmentation to address some of the issues with non-representativity of the data by re-sampling in time to match the cadence of the test data, and adding noise to the training data that was similar to the (larger) scatter in the test data, as summarized in Table~\ref{tab:methods}. 

\textbf{Class imbalance in \acro~--}
In addition to the test data being non-representative of the training data set, the classes within the training and test data sets had significantly different proportions relative to the total population of objects. This class imbalance reflects true imbalances in nature, and was hence a designed feature of \acro. Successful classifiers accounted for these imbalances by oversampling rare light curves, augmenting fluxes and errors or including test data with pseudo labels. The approaches employed by the various classifiers are shown in Table~\ref{tab:methods}.

\section{Conclusion}
\label{sec:discussion/conclusion}

The \acro~studies evaluate the classification probabilities derived by a range of methods for classifying a heterogeneous assortment of astronomical transients and variables. While previous challenges and data sets have focused on classifier performance related to one type of object or science case, \acro~represents the most complete simulation of future photometric surveys to date and challenged the community to take into account degeneracies between different types of interesting objects without a unifying science application and to prepare for future data that will contain novel/unknown objects that require classification and flagging for observational follow-up.  

Accurate classification of transients in future surveys remains one of the key challenges for time-domain observational astrophysics. While astronomers with domain knowledge are able to develop classification/selection methods tailored to individual science cases, the creativity and flexibility of the data science entries to \acro~suggests a collaborative future across disciplines will be key to making progress.

The \acro~challenge yielded a few key insights:
\begin{itemize}
    \item data augmentation of sparse non-representative training data are key to accurate performance on the full test set;
    \item a metric weighted by the number of objects in a class ensures that classification performance on the most populous class does not dominate overall performance;
    \item data fitting and feature selection are the most computationally time-consuming parts of any classification challenge;
    \item one successful approach is pulling out a larger set of features over which to train, and then determining the feature importance before classifying the objects on the smaller subset of features;
    \item objects with similar light curve shapes are broadly degenerate (e.g. SNIax and SNIa) and may present challenges to future surveys, motivating the inclusion of additional information about the host galaxy, where applicable;
    \item no strong preference for any one machine learning architecture was found; successful classification procedures often combine a range of methods from neural networks to tree-based approaches;
    \item domain knowledge helped mainly in identifying more useful features over which to train, which was an asset, given that this was often the most time-consuming part of the classification;
    \item all classifiers struggled to classify objects with low frequency in the training data, or novel (unknown) objects
\end{itemize}
Classifying objects from their photometric information alone will unlock the potential of using these interesting objects in a range of science applications. 

This challenge assumed that the full light curves for the objects were available. Long-lived transients can be followed up with ground-based resources within a few days of detection in a photometric survey. For some of the most interesting objects, however, waiting even a few days will mean losing valuable information about the evolution of the transient. As such, early classification of these bright objects becomes essential. We did not focus on the problem of early classification in \acro, but the next step in using these simulations to prepare for \lsst~will be to provide an extra performance score to classifiers capable of rapidly classifying transient sources, based on both their light curves and any additional data about their environment and location. This may prove more challenging or prone to bias than classification using the full light curve.

\subsection*{Acknowledgments}
This paper has undergone internal review in the \lsst~Dark Energy Science Collaboration. 

Author contributions are listed below. \\
R.~Hlo\v{z}ek: data curation, formal analysis, funding acquisition, investigation, project administration, software, supervision, validation, visualization, writing - editing, lead writing - original draft \\
K.~Ponder: software development and implementation for validation, testing, visualization and analysis of results, writing - editing, writing - original draft \\
A.I.~Malz: conceptualization, formal analysis, investigation, methodology, visualization, writing - editing, writing - original draft \\
M.~Dai:  software development and implementation for validation, testing, analysis of results, writing - editing \\
G.~Narayan: simulation, software, example kit, validation, visualization, writing - editing, writing - original draft \\
E.E.O.~Ishida: software development and implementation for validation, testing, Kaggle liason, analysis of results, writing - editing \\
T.~Allam Jr: investigation, development of software for simulation validation\\
A.~Bahmanyar:  investigation, implementation of metric methodology and software\\
R.~Biswas: conceptualization, methodology, software for validation and testing, writing - editing  \\
L.~Galbany: implementation of validation software, writing - editing \\
S.W.~Jha:  conceptualization, simulation validation \\
D.~Jones:   conceptualization, simulation validation, writing - editing\\
R.~Kessler: simulation development and validation testing, data generation, writing - editing \\
M.~Lochner:  conceptualization, simulation validation, writing - editing \\
A.A.~Mahabal:   conceptualization, simulation validation\\
K.S.~Mandel:   conceptualization, simulation validation, writing - editing\\
J.R.~Mart\'inez-Galarza:  conceptualization, simulation validation, writing - editing \\
J.D.~McEwen:   conceptualization, simulation validation\\
D.~Muthukrishna:   conceptualization, simulation validation, results analysis and testing, writing - editing\\
H.V.~Peiris: conceptualization, supervision, funding acquisition, writing - editing  \\
C.M.~Peters:   conceptualization, simulation validation \\
C.N.~Setzer:  conceptualization, simulation validation \\



The {\PLASTICC} data set generation relied on numerous members of the astronomical community to provide models of astronomical transients and variables. 
{\PLASTICC} is made possible through an inter-collaboration agreement between LSST's Dark Energy Science Collaboration and the Transient and Variable Stars Science Collaboration. These models are described in \cite{kessler/etal:2019_PLAsTiCC}.  
 We are grateful to Federica Bianco and staff at NYU for hosting the PLAsTiCC team meeting. 
The awards for PLAsTiCC contest were generously supported by Kaggle. 
We thank staff at Kaggle, particularly Elizabeth Park, Maggie Demkin and Sohier Dane, for their help making the contest a success. 
We thank Federica Bianco, Francisco F\"{o}rster Bur\'{o}n, Chad Schafer and Melissa Graham for providing internal review of this document for the DESC and TVS collaborations, and thank Seth Digel for comments on this draft.  This work was supported by an LSST Corporation Enabling Science grant. 
The financial assistance of the National Research Foundation (NRF) towards this research is hereby acknowledged. 
Opinions expressed and conclusions arrived at, are those of the authors and are not necessarily to be attributed to the NRF. 
UK authors acknowledge funding from the Science and Technology Funding Council. 
Canadian co-authors acknowledge support from the Natural Sciences and Engineering Research Council of Canada. 
The Dunlap Institute is funded through an endowment established by the David Dunlap family and the University of Toronto. 
The authors at the University of Toronto acknowledge that the land on which the University of Toronto is built is the traditional territory of the Haudenosaunee, and most recently, the territory of the Mississaugas of the New Credit First Nation. 
They are grateful to have the opportunity to work in the community, on this territory.
R.~H. acknowledges support from CIFAR, and the Azrieli  and Alfred. P. Sloan foundations. 
K.~A~.P. acknowledges support from the Berkeley Center for Cosmological Physics; the Director, Office of Science, Office of High Energy Physics of the U.S. Department of Energy under Contract No. DE-AC02-05CH11231; and U.S. Department of Energy Office of Science under Contract No.DE-AC02-76SF00515.
E.~E.~O.~I. acknowledges support from CNRS 2017 MOMENTUM grant.
A.~I.~M acknowledges support from the Max Planck Society and the Alexander von Humboldt Foundation in the framework of the Max Planck-Humboldt Research Award endowed by the Federal Ministry of Education and Research.
L.~G. was funded by the European Union's Horizon 2020 research and innovation programme under the Marie Sk\l{}odowska-Curie grant agreement No. 839090. 
D.~O.~J acknowledges support provided by NASA Hubble Fellowship grant HST-HF2-51462.001 awarded by the Space Telescope Science Institute, which is operated by the Association of Universities for Research in Astronomy, Inc., for NASA, under contract NAS5-26555.  D.O.J. also acknowledges support from the Gordon \& Betty Moore Foundation.
This work has been partially supported by the Spanish grant PGC2018-095317-B-C21 within the European Funds for Regional Development (FEDER).  
C.~N.~S., R.~B. and H.~V.~P. were partially supported by the research environment grant “Gravitational Radiation and Electromagnetic Astrophysical Transients (GREAT)" funded by the Swedish Research Council (VR) under Dnr 2016-06012. RB and HVP were additionally supported by the research project grant “Understanding the Dynamic Universe” funded by the Knut and Alice Wallenberg Foundation under Dnr KAW 2018.0067.
M.~D. acknowledges support from the Horizon Fellowship at the Johns Hopkins University.
At Rutgers University (M.~D., S.~W.~J.), this research was supported by NSF award AST-1615455 and DOE award DE-SC0011636. The DESC acknowledges ongoing support from the Institut National de Physique Nucl\'eaire et de Physique des Particules in France; the  Science \& Technology Facilities Council in the United Kingdom; and the Department of Energy, the National Science Foundation, and the LSST Corporation in the United States.  DESC uses resources of the IN2P3 Computing Center (CC-IN2P3--Lyon/Villeurbanne - France) funded by the Centre National de la Recherche Scientifique; the National Energy Research Scientific Computing Center, a DOE Office of Science User  Facility supported by the Office of Science of the U.S.\ Department of Energy under Contract No.\ DE-AC02-05CH11231; STFC DiRAC HPC Facilities,  funded by UK BIS National E-infrastructure capital grants; and the UK 
particle physics grid, supported by the GridPP Collaboration.  This work was performed in part under DOE Contract DE-AC02-76SF00515. 




\bibliographystyle{apsrev}
\bibliography{results,metric}


\end{document}